\documentclass[%
 rmp,
 amsmath,amssymb,
 aps,
]{revtex4-2}
\usepackage{natbib}
\usepackage[section]{placeins}
\usepackage{booktabs,threeparttable}

\usepackage{graphicx}
\usepackage{dcolumn}
\usepackage{bm}
\usepackage{float}
\usepackage{caption}
\usepackage{calrsfs}
\DeclareMathAlphabet{\pazocal}{OMS}{zplm}{m}{n}

\usepackage{amsthm}

\theoremstyle{definition}

\usepackage{enumitem}
\usepackage{eurosym}
\usepackage{hyperref}
\usepackage{bbold}

\usepackage[
scale=0.7, marginratio={1:1, 2:3}
]{geometry}

\usepackage{color}

\begin{document}
\title{
Equity auction dynamics: latent liquidity models with activity acceleration 
}

\author{Mohammed Salek}%
 \email{mohammed.salek@centralesupelec.fr}
\author{Damien Challet}%
 \email{damien.challet@centralesupelec.fr}
\author{Ioane Muni Toke}%
 \email{ioane.muni-toke@centralesupelec.fr}

\affiliation{Université Paris-Saclay, CentraleSupélec,  Laboratoire de Mathématiques et Informatique pour la Complexité et les Systèmes, 
  91192 Gif-sur-Yvette, France}%

\date{\today}

\begin{abstract}

Equity auctions display several distinctive characteristics in contrast to continuous trading. As the auction time approaches, the rate of events accelerates causing a substantial liquidity buildup around the indicative price. This, in turn, results in a reduced price impact and decreased volatility of the indicative price. In this study, we adapt the latent/revealed order book framework to the specifics of equity auctions. We provide precise measurements of the model parameters, including order submissions, cancellations, and diffusion rates. Our setup allows us to describe the full dynamics of the average order book during closing auctions in Euronext Paris. These findings support the relevance of the latent liquidity framework in describing limit order book dynamics. Lastly, we analyze the factors contributing to a sub-diffusive indicative price and demonstrate the absence of indicative price predictability.

\end{abstract}

\maketitle

\tableofcontents 

\parskip 0.1cm
\section*{Introduction}

Auctions play an essential role in modern equity markets, facilitating the matching of large volumes at a single price. Major equity exchanges such as Euronext (Paris), Xetra (Germany), and LSE (UK) among others, start and end trading days with an auction to set opening and closing prices for liquid stocks. The matched volume in the closing auction is a significant portion of the daily exchanged volume. This closing volume has seen consistent growth, particularly in European markets, where it can surpass half the daily volume on days of index rebalancing and derivative expiry \cite{raillon2020growing}. To enhance liquidity during midday, typically the period of lowest liquidity, and to mitigate the exposure to high-frequency market makers which dominate the total turnover during day trading \cite{AMF2017}, some primary exchanges such as Xetra and LSE, also introduced a daily intraday auction. For less liquid stocks, \citet{euronext_faq} implements the double-fixing trading method, i.e., two auctions per day and no continuous double auction. Some researchers advocate for periodic batch auctions as an alternative market design to continuous trading \cite{paul2021optimal,derchu2020ahead,budish2015high}. In 2015, \citet{CBOE_PA} implemented periodic batch auctions for liquid European stocks. 

During the so-called auction accumulation period, auction limit and market orders can be sent, modified, or canceled but no transaction occurs. At all times, an indicative price can be computed: it maximizes the matched volume (indicative volume) and minimizes the remaining order imbalance at the indicative price (surplus). Depending on the exchange, the order book data can be fully open: opening and closing auctions in the \citet{LSE_faq}, or partially opaque: the exchange only disseminates the indicative price, volume, the surplus, and its side \cite{Xetra_faq}. At the auction time, the exchange clears all matched orders at the auction price. Unlike US equity auctions where continuous trading runs in parallel, continuous trading is halted during accumulation periods in European equity markets.

Auctions are found empirically to improve the overall price formation process \cite{pagano2003closing}, even when the market is fragmented \cite{boussetta2017role}. Similarly, \citet{euronext2021} find that the cost of trading during closing auctions in European markets is reduced by a factor two compared to open markets. \citet{challet2018dynamical} examine US opening and closing auctions and find markedly different auction response functions for each auction. \citet{challet2019strategic} shows that as the auction clearing approaches, the indicative price volatility decreases while the rate of events accelerates. \citet{salek2023price} analyze equity auctions on Euronext Paris and find that price impact at the auction time is first zero due to large limit orders that are present at the auction price, then linear for most days.

While there is an extensive literature on the microstructure modeling of open markets \cite{chakraborti2011econophysics1,chakraborti2011econophysics2,bouchaud2018trades,lehalle2018market}, auction-specific models are relatively rare \cite{derksen2020clearing,mendelson1982market,toke2015exact}. The zero-intelligence model of \citet{donier2016walras} is a promising framework for Walrasian auctions; this model assumes the existence of a latent (hidden) limit order book containing all buy and sell intentions that may be partially revealed in the visible limit order book, building on \citet{toth2011anomalous}. Remarkably, the latent order book model is able to reproduce the shape of market impact with minimal ingredients.

Latent order book models have been the subject of extensive research \cite{donier2015fully,lemhadri2019price,Benzaquen2018,benzaquen2018fractional,mastromatteo2014agent}. Recently, \citet{dall2019does} introduced a conversion mechanism between the latent and revealed order books and fitted the resulting model to market data convincingly. In this work, we specifically adapt the framework of \citet{dall2019does} to equity auctions, allowing for more generic dependencies of the model parameters on price and time. In addition to stationary solutions, we demonstrate that general solutions can be obtained in closed form when diffusion is negligible. When diffusion is not negligible, numerical schemes are used to fit the full auction dynamics. Thanks to high-quality tick-by-tick data, we meticulously measure the price and time dependencies of the submission, cancellation, and diffusion rates within our framework. We show that the time acceleration near the clearing can be achieved using assumptions similar to those introduced by \citet{alfi2009people,alfi2007conference} which describe human behavior when faced with a deadline.

Our main findings are as follows:
\begin{enumerate}
    \item Calibrating the average book at auction time under our framework requires only a few essential parameters. This provides estimates of the involved latent liquidity.
    \item The submission rate is found to be an exponentially decreasing function of the distance to the indicative price, and inversely proportional to the remaining time to the auction clearing, both close enough to the indicative price.
    \item The cancellation rate is predominantly influenced by high-frequency agents. It is a decreasing time function at the start of the auction, then inversely proportional to the remaining time to the auction clearing around the indicative price.
    \item The revealed diffusion coefficient is primarily driven by the indicative price volatility, with price reassessments being, on average, negligible.
    \item Despite the indicative price being sub-diffusive, it is nevertheless efficient in the sense of \cite{chen2017anomalous}.
\end{enumerate}

\section{Modeling the auction book}
\label{sec:auction_model}

\subsection{Model description}

Our starting point is the coupled reaction-diffusion equations for the revealed $\rho^{(r)}$ and the latent $\rho^{(l)}$ orders densities derived in \citet{dall2019does}: they posit the existence of interactions between the latent order book $\rho^{(l)}$ and the revealed order book $\rho^{(r)}$. The reveal rate $\nu_r\Gamma_r$ is the rate at which latent trading intentions in the latent order book materialize into actual orders in the revealed order book and the \textit{unreveal} rate $\nu_l\Gamma_l$ is the rate at which actual orders revert back to latent intentions. In addition, a diffusion mechanism is included in both the revealed and the latent order book corresponding to price updates in each order book with diffusion coefficient $D_r$ and $D_l$, respectively. 

Assuming the absence of exogenous depositions and cancellations in the latent order book, an order submission is equivalent to an order revelation, and similarly, an order cancellation is equivalent to an order unrevelation. By allowing for general price and time dependencies, the buy side equations read
\begin{equation}
\begin{cases}
    \partial_t \rho_{B}^{(r)} &= D_r \partial_{xx} \rho_{B}^{(r)} + (\nu_r\Gamma_r)(-x,t)\rho_{B}^{(l)} - (\nu_l\Gamma_l)(-x,t)\rho_{B}^{(r)};\\
    \partial_t \rho_{B}^{(l)} &= D_l \partial_{xx} \rho_{B}^{(l)} - (\nu_r\Gamma_r)(-x,t)\rho_{B}^{(l)} + (\nu_l\Gamma_l)(-x,t)\rho_{B}^{(r)},
\end{cases}
\label{eq:modelB}
\end{equation}
where $x = \log(p/p_t)$ is the centered log price around the log indicative price $\log(p_t)$. 
For the sell side, we have 
\begin{equation}
\begin{cases}
    \partial_t \rho_{S}^{(r)} &= D_r \partial_{xx} \rho_{S}^{(r)} + (\nu_r\Gamma_r)(x,t)\rho_{S}^{(l)} - (\nu_l\Gamma_l)(x,t)\rho_{S}^{(r)};\\
    \partial_t \rho_{S}^{(l)} &= D_l \partial_{xx} \rho_{S}^{(l)} - (\nu_r\Gamma_r)(x,t)\rho_{S}^{(l)} + (\nu_l\Gamma_l)(x,t)\rho_{S}^{(r)}.
\end{cases}
\label{eq:modelS}
\end{equation}

We complement Eqs \eqref{eq:modelB} and \eqref{eq:modelS} with a set of boundary conditions. Let us focus on the sell side for the time being. It makes sense to assume that the latent liquidity book $\rho_{S}^{(l)}$ is an increasing function of $x$, as more people are willing to sell assets at a higher price. This means that one can impose a boundary condition on the slope of the latent order book:  $\partial_x \rho_{S}^{(l)} \underset{x \rightarrow + \infty}{\longrightarrow} a >0 $, which corresponds to the latent liquidity parameter of \citet{donier2015fully}, and that $\rho_{S}^{(l)}$ does not diverge when $x \rightarrow -\infty$, which reflects the fact that the number of people willing to sell at a vanishing price is not infinite. We also impose that the revealed order book $\rho_S^{(r)}$ does not diverge for large prices $|x| \rightarrow + \infty $. This means that agents tend not to reveal their reservation price when it is far away from the indicative price. In practice, agents can send market orders or matchable limit orders far away from the indicative price in order to guarantee their participation in the auction volume. Market orders are not included in the densities $\rho_S^{(r)}$ and $\rho_B^{(r)}$ as latent liquidity models are defined in the reference frame of the indicative price.

Similar boundary conditions hold for the buy side.

In the original framework, $\nu_l=\nu_r$ is a constant rate, and $\Gamma_r$ (resp. $\Gamma_l$) is conceived as a probability function of the relative price $x$ for revealing a latent intention (resp. unrevealing a public intention), with $0\leq \Gamma \leq 1$. In the context of equity auctions, the remaining time to the auction clearing plays an essential role \cite{challet2019strategic}. Thus, we allow for the quantities $\nu_r\Gamma_r$ and $\nu_l \Gamma_l$ to depend jointly on $x$ and $t$. However, we will often posit that the variable separation is possible $(\nu\Gamma)(x,t) = \nu(t)\Gamma(x)$ for the sake of analytical tractability and interpretability. 

In addition, the initial model comprises a reaction term that is formally written $\kappa R_{SB} = \kappa \rho_{S}^{(r)}\rho_{B}^{(r)}$, with $\kappa \rightarrow +\infty$ in order to make both densities interact and transactions happen; for sufficiently large $\kappa$, no overlap between the buy and sell densities is possible. In Walrasian auctions instead, no transaction takes place before the clearing, and buy and sell limit orders usually overlap. When the order book is partially opaque during the accumulation period, it is reasonable to believe that order densities interact solely through the knowledge of the indicative price. Therefore, when considered in the reference frame of the indicative price, buy and sell order densities should evolve independently, which leads us to set $\kappa = 0$.

Lastly, in open markets, the current price $p_t$ is the point where vanishing supply meets vanishing demand; accordingly, \citet{dall2019does} define $p_t$ as the point where $p \rightarrow \left(\rho_B^{(r)}-\rho_S^{(r)}\right)(p)$ changes sign. In auctions instead, the indicative price $p_t$ is determined by equalizing supply and demand 
\begin{equation}   
S(p_t) + MO_{S,t} = D(p_t) +MO_{B,t},
\label{eq:S+M=D+M}
\end{equation}
where $S(p) = \int_{p'<p} \rho_S^{(r)}(p') \mathrm{d}p' $, $D(p) = \int_{p'>p} \rho_B^{(r)}(p') \mathrm{d}p'$, and $MO_B$ and $MO_S$ is the volume of buy and sell auction market orders. Equation \eqref{eq:S+M=D+M} makes it clear that only the difference of market order volume has an influence on the indicative price. As latent liquidity models are defined in the reference frame of the indicative price, market orders are not included in the densities $\rho_S^{(r)}$ and $\rho_B^{(r)}$.

 When centering equations around $\log(p_t)$, we implicitly assume that the indicative price evolves independently of order densities. However, changes in the sell (resp. buy) order density for $x<0$ (resp. $x>0$) directly influence supply (resp. demand) subsequently changing the indicative price. We disregard this effect and consider that the dynamics of the indicative price are independent of average order densities around the indicative price. In the following, we focus on the sell-side equations \eqref{eq:modelS}, and denote $\rho_S$ as $\rho$ when there is no ambiguity.
 
\subsection{A stationary solution}

The simplest stationary solution of Eqs \eqref{eq:modelS} is obtained when both $\nu_l\Gamma_l$ and $\nu_r\Gamma_r$ are time-independent and there is no diffusion in the revealed order book ($D_r =0$), while orders may diffuse in the latent order book ($D_l \ge0$). 
Eqs \eqref{eq:modelS} reduce to
\begin{equation}
\begin{cases}
    0 &= \nu_r\Gamma_r\rho^{(l)} - \nu_l\Gamma_l\rho^{(r)},\\
    0 &= D_l \partial_{xx} \rho^{(l)} - \nu_r\Gamma_r\rho^{(l)} + \nu_l\Gamma_l\rho^{(r)},
\end{cases}
\end{equation}
which further simplifies to
\begin{equation}
\label{eq:rev_stationnary}
\begin{aligned}
    \rho^{(r)} &= \frac{\nu_r}{\nu_l} \cdot \frac{\Gamma_r}{\Gamma_l} \cdot \rho^{(l)}\\
    D_l\partial_{xx} \rho^{(l)} &= 0.  
\end{aligned}
\end{equation}
We solve Eq. \eqref{eq:rev_stationnary} separately on $\mathbb{R}^+$ and $\mathbb{R}^-$ as the first derivative of $x \rightarrow \Gamma_{r/l}(x)$ might not be continuous at $x=0$. Thus, whenever $D_l \ne 0$, the latent order book should be linear. Incorporating the boundary conditions $\partial_x \rho^{(l)} \underset{x \rightarrow + \infty}{\longrightarrow} a >0 $ and $\rho^{(l)} \underset{x \rightarrow - \infty}{\longrightarrow} b \geq 0 $ yields
\begin{equation}
        \rho^{(l)}(x) = \max(ax+b,b).
\label{eq:dxx_null}
\end{equation}
Subsequently, the stationary revealed order book is 
\begin{equation}
        \rho^{(r)}(x) = \frac{\nu_r}{\nu_l} \cdot \frac{\Gamma_r(x)}{\Gamma_l(x)} \cdot \max(ax+b,b).
\label{eq:rho_r_st}
\end{equation}
As the revealed order book does not diverge for large prices, Eq. \eqref{eq:rho_r_st} imposes that $\Gamma_r$ decays faster than $\Gamma_l$ for $x \rightarrow - \infty$, and that $x \Gamma_r$ decays faster than $\Gamma_l$ for $x \rightarrow + \infty$. For instance, this condition is satisfied by an exponentially decaying submission rate and a constant cancellation rate.

\subsection{Dynamic solutions}

Here, we derive non-stationary solutions of Eqs \eqref{eq:modelS} in several cases. First, we suppose that  diffusion is negligible in both latent and revealed order books ($D_r = D_l = 0$). This is a sound approximation as the calibrated orders of magnitude of $D_l$ and $D_r$ do not significantly influence the order book shape (see section \ref{sec:num_solving}).

Summing Eqs \eqref{eq:modelS} yields 
\begin{equation}
    \partial_t (\rho^{(l)}+\rho^{(r)}) = 0,
    \label{eq:sum_st}
\end{equation}
which suggests defining the total density $\rho^{\Sigma} = \rho^{(l)}(x,t)+\rho^{(r)}(x,t)$.
Thus, Eq. \eqref{eq:sum_st} implies that 
\begin{equation}
    \rho^{\Sigma}(x,t) = \rho^{\Sigma}(x,t=0).
\end{equation}
Now, we posit the initial condition $\rho^{(r)}(x,t=0) = 0$, which is a reasonable approximation because the revealed order book at the beginning of the auction is negligible in comparison with the final auction book. Additionally, we set $\rho^{(l)}(x,t=0) = \max(ax+b,b)$ to satisfy the latent order book boundary conditions. Making these substitutions, we find that Eqs \eqref{eq:modelS} can be decoupled into
\begin{align}
    \partial_t \rho^{(r)} + \left(\nu_r\Gamma_r+\nu_l\Gamma_l\right)\cdot\rho^{(r)} &= \nu_r\Gamma_r\rho^{\Sigma};\\
    \partial_t \rho^{(l)} + \left(\nu_r\Gamma_r+\nu_l\Gamma_l\right)\cdot\rho^{(l)} &= \nu_l\Gamma_l\rho^{\Sigma},
\label{eq:decoupled_sys}
\end{align}
where  $\rho^{\Sigma} = \max(ax+b,b)$. 

\subsubsection{Time-independent rates}

\label{sec:simp_dynamic}

When $\nu_l\Gamma_l$ and $\nu_r\Gamma_r$ do not exhibit temporal dependencies, i.e., $(\nu \Gamma)(x,t) = \nu \cdot \Gamma(x)$,  Eqs \eqref{eq:decoupled_sys} yields the following expression for the revealed order book
\begin{equation}
    \rho^{(r)}(x,t) = \rho_{\infty} \cdot \left[1-e^{-\left(\nu_r \Gamma_r+\nu_l \Gamma_l\right) \cdot t}\right],
    \label{eq:sol_rates_stationary}
\end{equation}
where $ \rho_{\infty} =   (\nu_r \Gamma_r \cdot \rho^{\Sigma}) /(\nu_r \Gamma_r+\nu_l \Gamma_l)$. The obtained solution converges to $\rho_{\infty}$ in the long run, i.e., as \mbox{$t \longrightarrow +\infty$}. This solution replicates the dynamics of the revealed order book during auction phases when rates $\nu_l\Gamma_l$ and $\nu_r\Gamma_r$ can be considered time-independent. Moreover, it is valid for order book portions that are already in a stationary state. Finally when $(\nu_r \Gamma_r+\nu_l \Gamma_l) \cdot t \ll 1$, $\rho^{(r)}$ no longer depends on $\nu_l\Gamma_l$ anymore. This occurs for large values of $|x|$ where $\nu_r\Gamma_r + \nu_l\Gamma_l$ goes to zero while $t$ remains bounded
\begin{equation}
\begin{aligned}
         \rho^{(r)}(x,t) \underset{|x|\gg1}{\sim} \nu_r \cdot \Gamma_r(x) \cdot  \rho^{\Sigma}(x)\cdot t.
\end{aligned}
\label{eq:rho_large_x}
\end{equation}
Eq. \eqref{eq:rho_large_x} demonstrates that $\rho^{(r)}$ does not diverge when $|x| \rightarrow +\infty$ even when $\rho_{\infty}$ does, e.g. when $\Gamma_l$ decays faster than $\Gamma_r$. Note that this simple framework can not reproduce the accelerating auction dynamics as the clearing approaches \cite{challet2019strategic}. 

\subsubsection{Time-dependent rates}
\label{sec:mod_theo_deadline}

To replicate the accelerating order book activity as the auction time approaches, which usually results in a convex $\rho^{(r)}$ with respect to $t$ as $t \longrightarrow T$, we need to introduce a pressure from the auction deadline $T$. The auction deadline is the final time that ensures trading with certainty. For liquid stocks listed in Euronext, $T=$ 17:35:00 for the closing auction, and 09:00:00 for the opening auction. The clearing randomly occurs in a thirty-second window starting at $T$.\footnote{In the 28th of September 2015, Euronext introduced a random clearing window of thirty-second length for its equity auctions. The clearing randomly happens between 09:00:00 and 09:00:30 for the opening auction and 17:35:00 and 17:35:30 for the closing auction. This prevents fast agents from using low latency to take advantage of slower agents.} 

Drawing inspiration from \citet{alfi2007conference, alfi2009people}, who argue that the probability of registering at a conference is inversely proportional to the remaining time to the registration deadline, we posit that the submission rate $\nu_r\Gamma_r$ (resp. the cancellation rate $\nu_l\Gamma_l$) is time-dependent for $t\geq t_r^{(0)}$ and constant for $t\leq t_r^{(0)}$ (resp. $t_l^{(0)}$), where $t_{r/l}^{(0)}$ is a cut-off time in $]0,T[$. More precisely, we set
\begin{equation}
    \begin{aligned}
        (\nu_r\Gamma_r)(x,t) &= \frac{C_r}{\gamma_r + T-t}\cdot \Gamma_r(x)  , \quad  &t\geq t_r^{(0)}; \\
        (\nu_l\Gamma_l)(x,t) &= \frac{C_l}{\gamma_l +T-t} \cdot \Gamma_l(x) , \quad &t\geq t_l^{(0)},
    \end{aligned}
    \label{eq:gamma_deadline}
\end{equation}
with $C_r,C_l>0$, $\gamma_l,\gamma_r\geq 0 $. A strictly positive $\gamma_r$ (resp. $\gamma_l$) indicates that the perceived deadline for submitting (resp. canceling) limit orders around the indicative price is $T+\gamma_r$ (resp. $T+\gamma_l$).

Assuming that $\gamma_r = \gamma_l = \gamma$ and $t_r^{(0)} = t_l^{(0)} = t^{(0)}$, we substitute $\nu_r\Gamma_r$ and $\nu_l\Gamma_l$ of Eqs \eqref{eq:gamma_deadline} into the first equation of \eqref{eq:decoupled_sys} and obtain for $t\geq t^{(0)}$
\begin{equation}
    \partial_t \rho^{(r)} + \frac{C_r \Gamma_r + C_l\Gamma_l}{\gamma+ T-t}\cdot\rho^{(r)} =  \frac{C_r\Gamma_r}{\gamma+ T-t} \cdot \rho^{\Sigma},
\label{eq:sol_deadline}
\end{equation}
whose solution is given by
\begin{equation}
     \rho^{(r)}(x,t) = \rho_T - (\rho_T-\rho_0)\cdot\left(\frac{\gamma + T-t}{\gamma + T - t^{(0)} }\right)^{C_r \Gamma_r + C_l\Gamma_l},
 \label{eq:sol_realistic}
\end{equation}
where $\rho_T = C_r\Gamma_r\rho^{\Sigma}/ (C_r\Gamma_r+C_l\Gamma_l)$, and $\rho_0$ is obtained by substituting $t = t^{(0)}$ in Eq. \eqref{eq:sol_rates_stationary}.

The functional shape of Eq. \eqref{eq:sol_realistic} is convex w.r.t. $t$ for all $0\leq t \leq T+ \gamma$ whenever $C_r \Gamma_r + C_l\Gamma_l<1$. It converges to a finite solution $\rho_T$ as  $t \longrightarrow \gamma + T$. Assuming that $\Gamma_l$ decays faster than $\Gamma_r$ yields a divergent $\rho_T$ for large $|x|$ and the solution of Eq. \eqref{eq:sol_realistic} becomes incompatible with the boundary conditions of the revealed order book. However, it should be noted that the time dependence of the reveal and unreveal rates in Eq. \eqref{eq:gamma_deadline} is only valid in a limited region around the indicative price and that for large values of $|x|$ the time dependency vanishes.

\begin{figure}
    \centering
    \includegraphics[scale = 0.295]{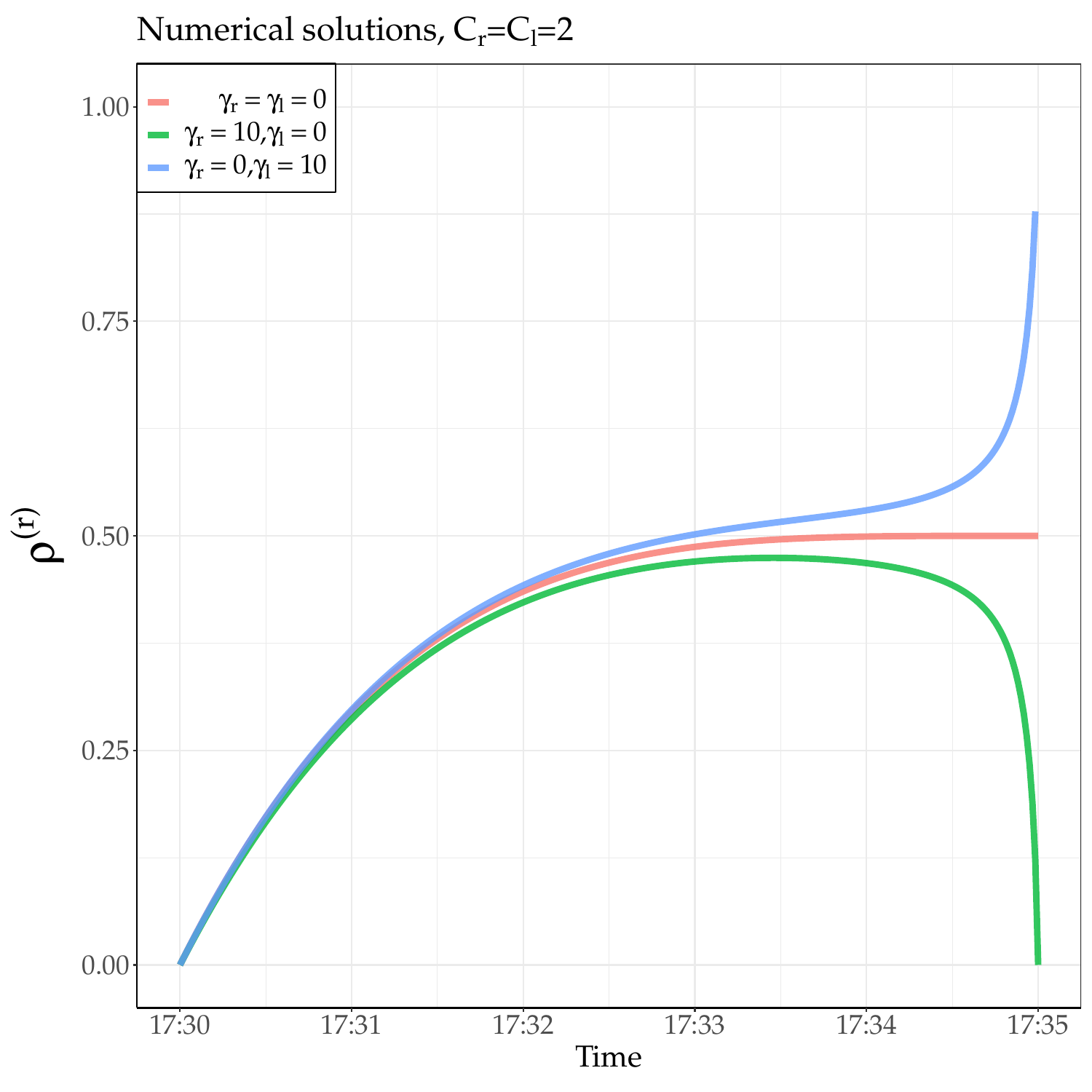}
    \includegraphics[scale = 0.295]{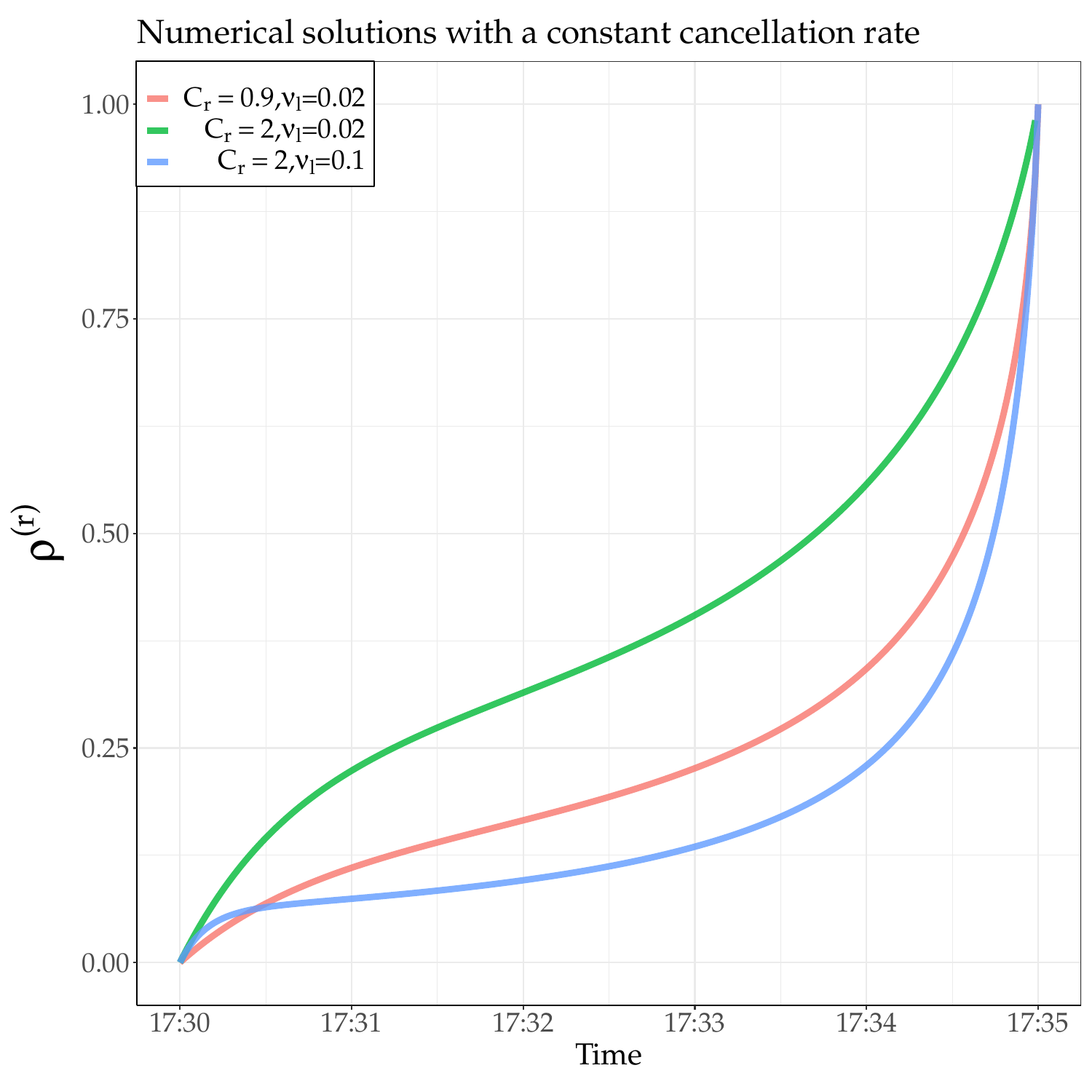}
    \caption{Left panel: numerical solutions of Eq. \eqref{eq:diff_gammas} for $C_r=C_l=2$. Right panel: numerical solutions of Eq. \eqref{eq:constant_nu_l} with a constant cancellation rate $\nu_l$.}
    \label{fig:cv_time_sol}
\end{figure}

Convex solutions of time around the auction deadline can be obtained even if $C_r, C_l>1$ provided that $\gamma_l>\gamma_r$, i.e., when the perceived deadline of cancellations occurs later than that of submissions. The ordinary differential equation verified by the revealed order book reads
    \begin{equation}
    \partial_t \rho^{(r)} + \left(\frac{C_l}{\gamma_l+T-t}+\frac{C_r}{\gamma_r+T-t}\right)\cdot\rho^{(r)} = \frac{C_r}{\gamma_r+T-t}.
    \label{eq:diff_gammas}
    \end{equation}
Likewise, when the rate of cancellations is constant $(\nu_l\Gamma_l)(x,t) = \nu_l >0$, the resulting order book dynamic is convex as $t \longrightarrow T$, and the ordinary differential equation reads
    \begin{equation}
    \partial_t \rho^{(r)} + \left(\nu_l+\frac{C_r}{\gamma_r+T-t}\right)\cdot\rho^{(r)} = \frac{C_r}{\gamma_r+T-t}.
    \label{eq:constant_nu_l}
\end{equation}
Eqs \eqref{eq:diff_gammas}, \eqref{eq:constant_nu_l} are challenging to solve analytically, and we present numerical solutions in Figure \ref{fig:cv_time_sol} by varying the parameters of interest.

\subsubsection{Diffusion}

The diffusion coefficient is the sum of two contributions: idiosyncratic price reassessments and reactions to changes in the indicative price. We explore the limiting case where latent price reassessments equal the revealed ones $D_r = D_l = D> 0$. Summing once again Eqs \eqref{eq:modelS}, we obtain
\begin{equation}
    \partial_{t} \left(\rho^{(r)} + \rho^{(l)}\right) = D \cdot \partial_{xx} \left(\rho^{(r)} + \rho^{(l)}\right).
    \label{eq:rho_sigma_pde}
\end{equation}
The solution $\rho^{\Sigma} = \rho^{(r)}+ \rho^{(l)}$ of Eq. \eqref{eq:rho_sigma_pde} is then
\begin{equation}
    \begin{aligned}
        \rho^{\Sigma}(x,t) &= \frac{1}{\sqrt{4\pi D t}} \int_{\mathbb{R}} \rho^{\Sigma}(y,t=0) \cdot  e^{-\frac{(x-y)^2}{4Dt}} \mathrm{d}y;\\
        \rho^{\Sigma}(x,t=0) &= \max(ax+b, b).
    \end{aligned}
\end{equation}
In this scenario, Eqs \eqref{eq:modelS} can be decoupled, and order densities $\rho^{(r)}$ and $\rho^{(l)}$ satisfy 
\begin{equation}
\begin{cases}
    \partial_t \rho^{(r)} &= D \partial_{xx} \rho^{(r)} - \left(\nu_r\Gamma_r+\nu_l\Gamma_l\right)\cdot\rho^{(r)} + \nu_r\Gamma_r\rho^{\Sigma};\\
    \partial_t \rho^{(l)} &= D \partial_{xx} \rho^{(l)} - \left(\nu_r\Gamma_r+\nu_l\Gamma_l\right)\cdot\rho^{(l)} + \nu_l\Gamma_l\rho^{\Sigma}.
\end{cases}
\label{eq:decoupled_sys_D}
\end{equation}
If we further assume that $\nu_r\Gamma_r+\nu_l\Gamma_l$ is constant w.r.t. $x$ and $t$, the revealed order density can be obtained in a closed-form formula \cite{donier2016walras}, and diffusion leads to non-trivial interactions depending on the shape of the source terms. Outside this specific case ($D_r=D_l$ and $\nu_r\Gamma_r+\nu_l\Gamma_l$ constant), it is challenging to obtain closed-form solutions and we solve the general Eqs \eqref{eq:modelS} numerically in section \ref{sec:num_solving}.

\section{Empirical observations and calibrations}
\label{sec:measures}

In this section, we confront the model presented in section \ref{sec:auction_model} to real auctions in Euronext Paris. Using high quality data from BEDOFIH, we process detailed tick-by-tick closing auction data for five active stocks (TotalEnergies, Sanofi, BNP Paribas, LVMH, Société Générale) in Euronext Paris between 2013 and 2017. First, we reconstruct order book snapshots at the auction time to calibrate the stationary solution of our model. Next, we leverage the level-3 tick-by-tick data to measure the submission, cancellation, and diffusion rates in the revealed order book. Lastly, we reconstruct 1-second successive snapshots to calibrate the full dynamics of the revealed order book during the auction.

\subsection{Fitting order books at auction time}
\label{sec:static_fits}

Using Eq. \eqref{eq:rho_r_st} we can fit the empirical order book at auction time $\rho^{(r)}(t=T)$. The empirical order book at auction time is obtained by averaging order book snapshots at the closing auction time across days. We choose the simplest functional forms for the stationary submission and cancellation rates, i.e., an exponentially decreasing revelation probability function $\Gamma_r(x) = e^{-|x|/x_r}$, $x_r>0$ and a constant unrevelation probability function $\Gamma_l = 1$. Substituting these into Eq \eqref{eq:rho_r_st}, we fit the empirical order density at auction time. The obtained fits suggest that accuracy is improved by allowing for more than one exponential term. Consequently, we use the following ansatz
\begin{equation}
    \rho^{(r)}(x) = \frac{\nu_r}{\nu_l} \cdot 
        \max(ax + b, b) \cdot \left[ w \cdot e^{-|x|/x_r} + 
       (1- w) \cdot e^{-|x|/(k\cdot x_r)} \right],
    \label{eq:static_fit_two_exp}
\end{equation}
where $0\leq w \leq 1$, $k \geq 1$. This can be interpreted as having two types of agents with different price scales. In a similar framework, \citet{Benzaquen2018} find that the typical price scale is proportional to the square root of the typical timescale. Thus, fast agents can be characterized as having a smaller price scale $x_r$ and slow agents as having a larger price scale $kx_r$.

Note that the parameters $a$ and $b$ in Eq. \eqref{eq:static_fit_two_exp} can only be determined up to a factor $\nu_r/\nu_l$. Prior to delving into fits, we independently measure $\nu_r/\nu_l$ as the ratio of the number of submissions to the number of cancellations, in the vicinity of $x=0$, seconds before the clearing.
\begin{table}
\caption{Median value of $\nu_r / \nu_l$ for $N=1266$ closing auctions of TotalEnergies between 2013 and 2017.}
\begin{ruledtabular}
\begin{tabular}{ccc}
 $\nu_r / \nu_l$ & Buy side & Sell side \\
\hline
last 1 second & 1.17 & 1.2 \\
last 10 seconds & 1.13 & 1.15  \\
last 30 seconds & 1.03 & 1.05  \\
\end{tabular}
\end{ruledtabular}
\label{tab:nu_r_over_nu_l_median}
\end{table}
We report in Table \ref{tab:nu_r_over_nu_l_median} the median values of this ratio in the last second, 10 seconds, and 30 seconds before 17:35:00 for each side. These results imply a median value around 1.

We now proceed to fitting buy and sell densities at auction time, as well as their breakdown by latency (HFT, MIX, and NON\footnote{A participant is considered a high-frequency trader (HFT) if he meets one of the two following conditions:
\begin{itemize}[noitemsep,topsep=0pt]
    \item The average lifetime of its canceled orders is less than the average lifetime of all orders in the book, and it has canceled at least 100,000 orders during the year.
    \item The participant must have canceled at least 500,000 orders with a lifetime of fewer than 0.1 seconds, and the top percentile of the lifetime of its canceled orders must be less than 500 microseconds.
\end{itemize}
An investment bank meeting one of these conditions is described as mixed-HFT (MIX). If a participant does not meet any of the above conditions, it is a non-HFT (NON).}) for each of the five studied stocks. For instance, fitting the HFT order book yields an estimate of the HFT latent book, i.e., the latent book that contains trading intentions of  HFT-flagged agents only. This breakdown will prove useful in section \ref{sec:mes_gamma_r} where we measure the contribution by agent category to the global submission rate.

We run each minimization procedure from 18 different initializations in order to avoid local minima and use ordinary least squares to obtain a set of optimal parameters $(\hat{a},\hat{b},\widehat{x_r},\hat{k}, \hat{w})$.
\begin{figure}
    \centering
    \includegraphics[scale=0.37]{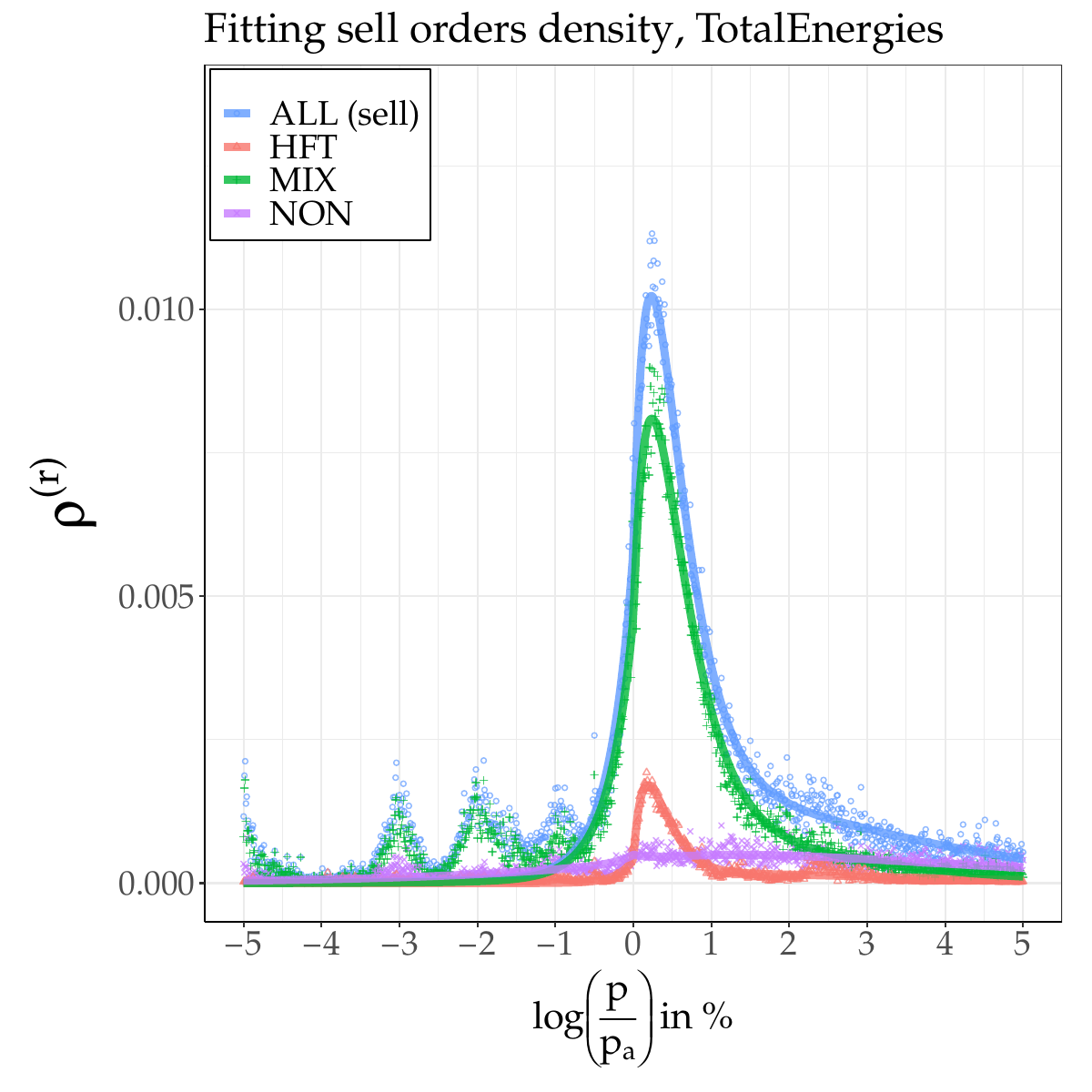}
    \includegraphics[scale=0.36]{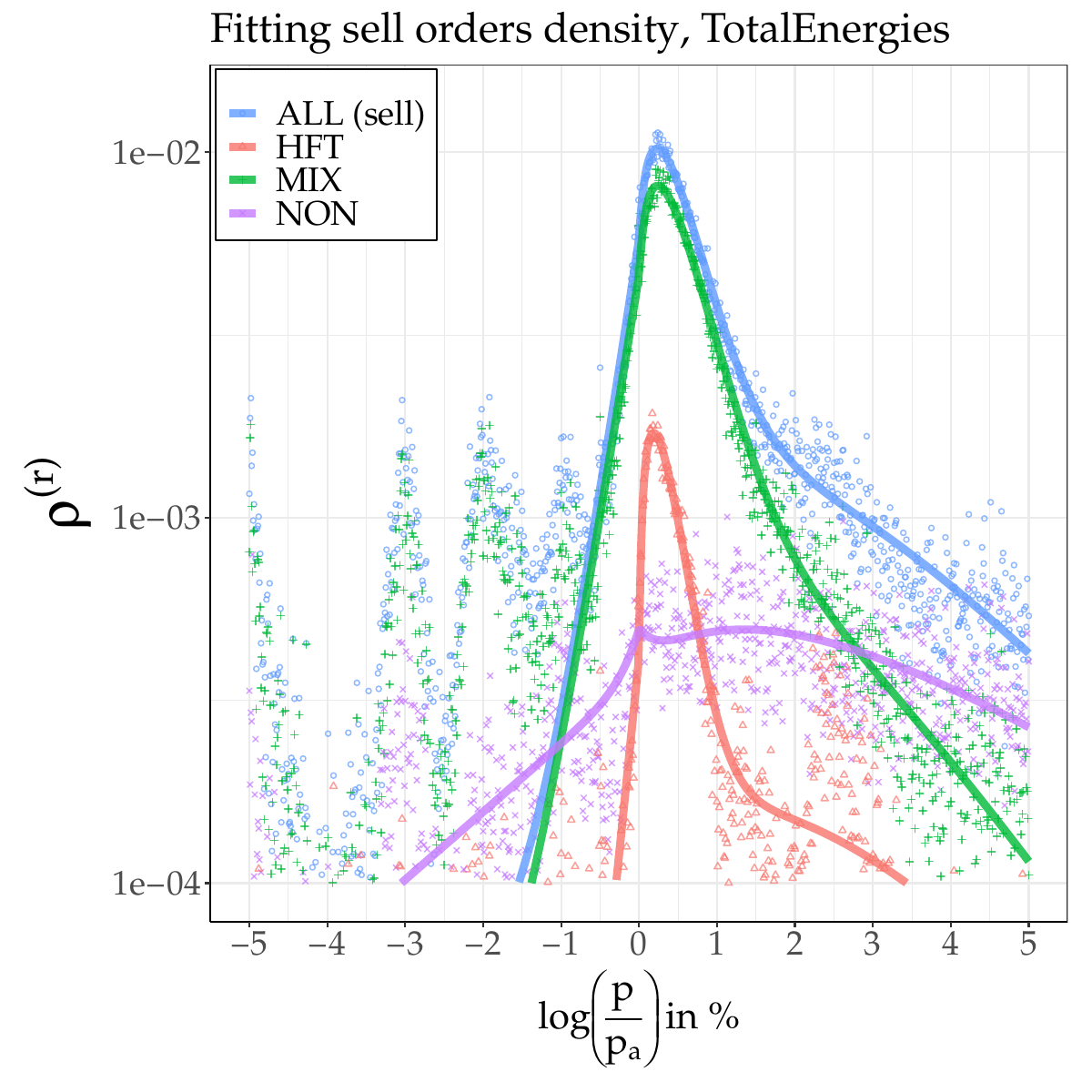}
    \caption{Fits (solid lines) with the functional of Eq. \eqref{eq:static_fit_two_exp} of the average orders' density $\rho^{(r)}$ just before the close clearing for TotalEnergies between 2013 and 2017. Left panel: ordinary scale. Right panel: Y-axis in log scale. The HFT flag denotes pure high-frequency traders, MIX denotes investment banks with high-frequency trading activities, and NON denotes traders without HFT activities.}
    \label{fig:Static_fits}
\end{figure}
We report the optimal estimates for each stock and side in Appendix \ref{sec:appendix_fits}, Table \ref{tab:static_fits_two_exp}, and present fits for TotalEnergies in Fig. \ref{fig:Static_fits}. Eq. \eqref{eq:static_fit_two_exp} provides accurate fits for $x>0$. However, it cannot reproduce the oscillations of liquidity at multiples of $-0.5\%$. These fluctuations result from punctual order submissions in these locations that undergo diffusion with coefficient $D_r$ seconds before the clearing.

\subsection{The empirical dynamics of the auction book}
\label{sec:measures_1}

\begin{figure}
    \centering
    \includegraphics[scale=0.295]{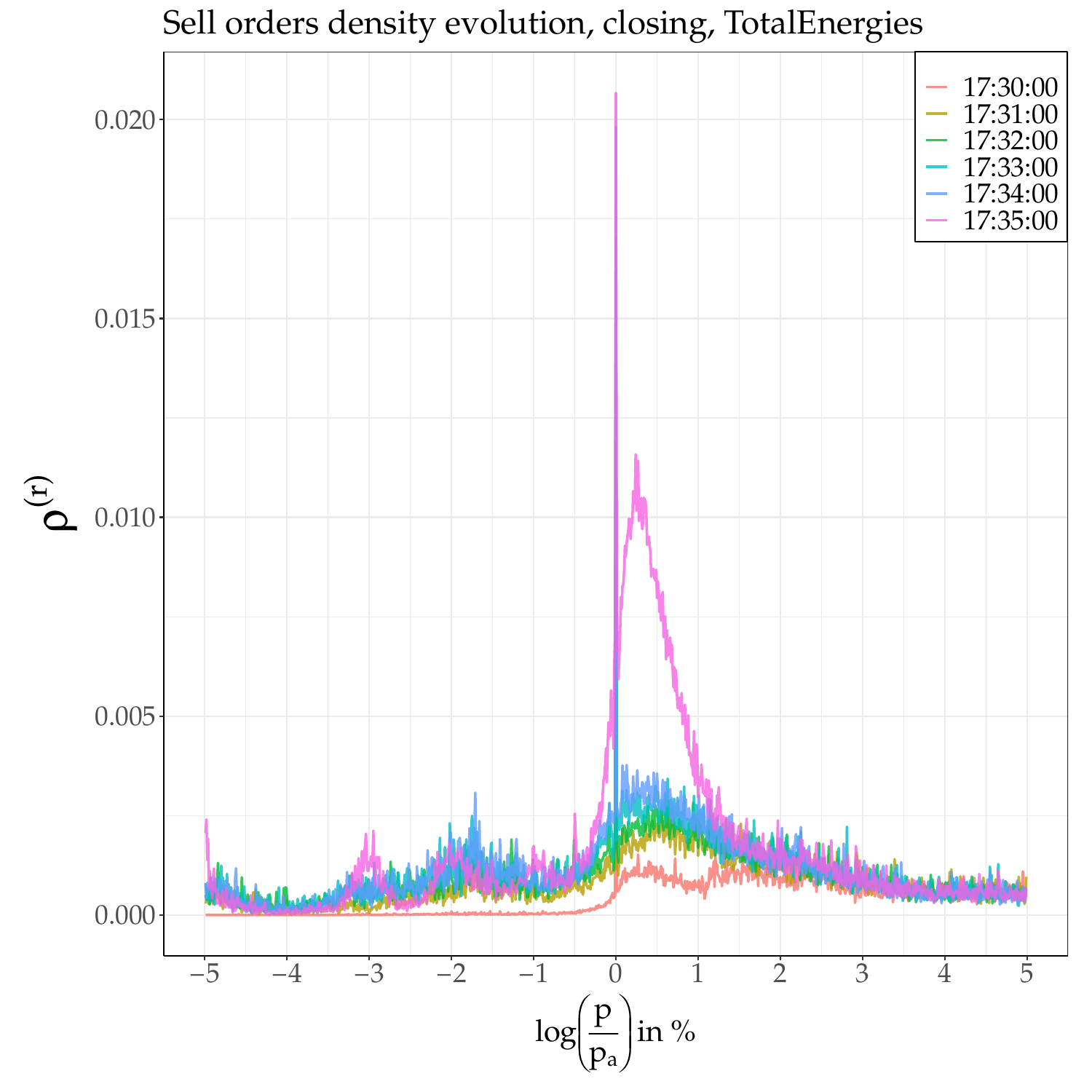}
    \includegraphics[scale=0.295]{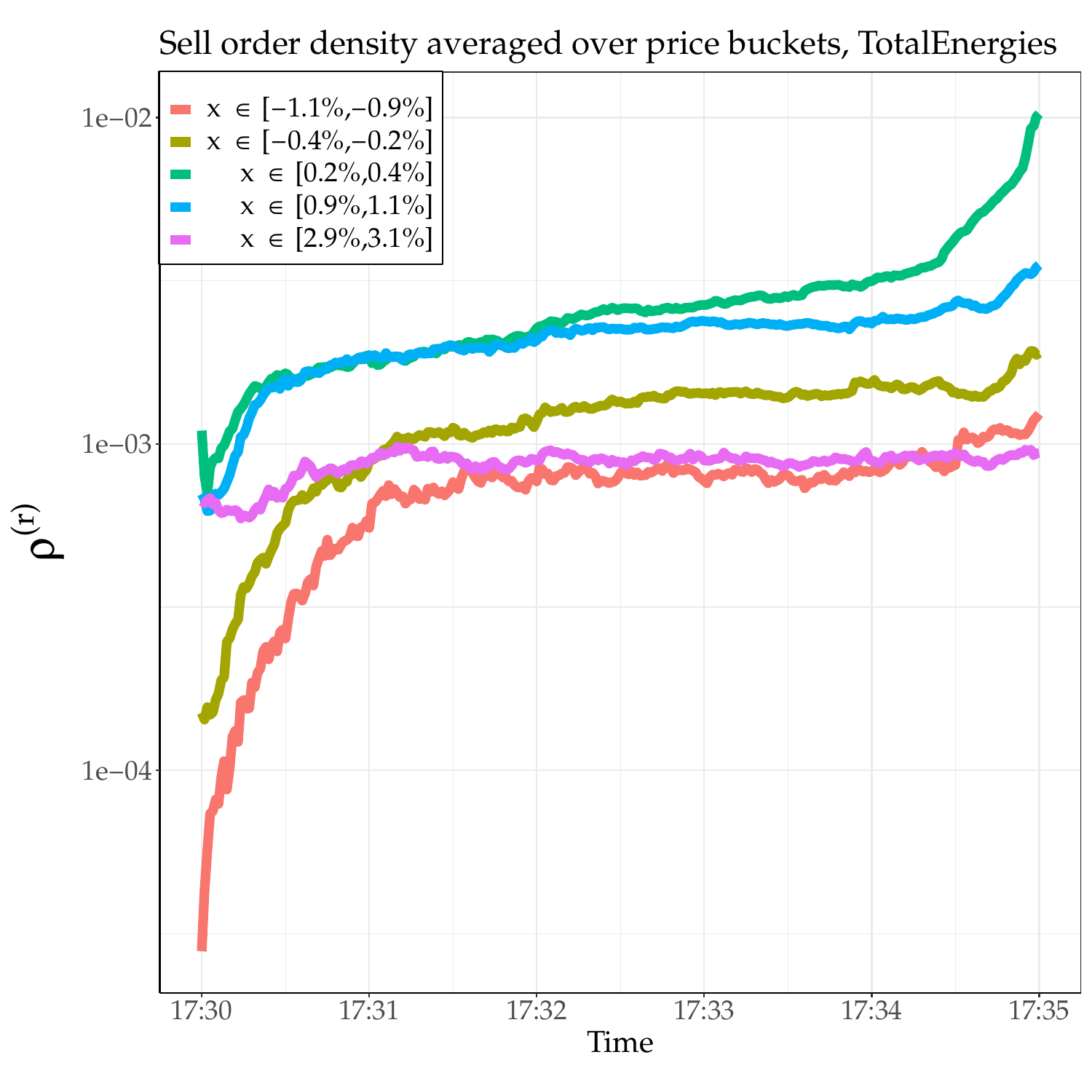}
    \caption{Empirical time and price dependencies of the average sell order density $\rho_S^{(r)}$ throughout the closing auction. We averaged auction book data for TotalEnergies.}
    \label{fig:LOB_evolution}
\end{figure}

Before providing measurements of the model parameters, we first present an overview of the dynamics of the empirical order density $(x,t) \rightarrow \rho^{(r)}(x,t)$. For that purpose, we perform 1-second order book snapshots during the closing auction averaged over days. We present in Fig. \ref{fig:LOB_evolution} the empirical functions $x \rightarrow \rho^{(r)}$ at round minutes and $t \rightarrow \rho^{(r)}$ at various prices. We observe that $\rho^{(r)}$ exhibits a skewed shape w.r.t. $x$, reaching its maximum for $x>0$ (which corresponds to non-matched orders for both buy and sell sides by convention). Its temporal dependency is initially concave, then accelerates towards the clearing. The time acceleration is more pronounced around the maximum argument and vanishes for large values of $|x|$.

Within the latent order book framework, the revealed order density $\rho^{(r)}$ evolves in time due to three mechanisms: submissions from latent order book $\rho^{(l)}$ with a rate $\nu_r\Gamma_r$, cancellations from the revealed book with a rate $\nu_l\Gamma_l$, and diffusion with a coefficient $D_r$. In this part, we provide empirical measurements of these rates. 

The reconstructed tick-by-tick data from BEDOFIH allows us to track changes in quantity for each price level. We view pure price updates without a change in quantity as a diffusion mechanism. Thus, we do not categorize pure price updates as cancellations from the corresponding previous price limits nor as submissions to the new ones.

\subsubsection{Measuring the cancellation rate: $\nu_l\Gamma_l$}
\label{sec:mes_gamma_l}

The cancellation rate $(\nu_l\Gamma_l)(p, t -\delta t \rightarrow t)$ for price $p$ between $t-\delta t $ and $t$ is defined as
\begin{equation}
    (\nu_l \Gamma_l)(p, t -\delta t \rightarrow t)= -\frac{1}{\delta t}\sum_{t-\delta t<t_i<t}
    \frac{\delta V(p,t_i)}{V_p(t_i)}\cdot \mathbb{1}_{\{\delta V(p,t_i)<0\}},
    \label{eq:estim_gamma_l}
\end{equation}
where $\delta V(p,t_i)$ is the volume change at limit price $p$ at time $t_i$ excluding volumes that diffused to another price $p'$, and $V_p(t_i)$ is the total volume at limit price $p$ at time $t_i^-$.

\begin{figure}
    \centering
    \includegraphics[scale=0.295]{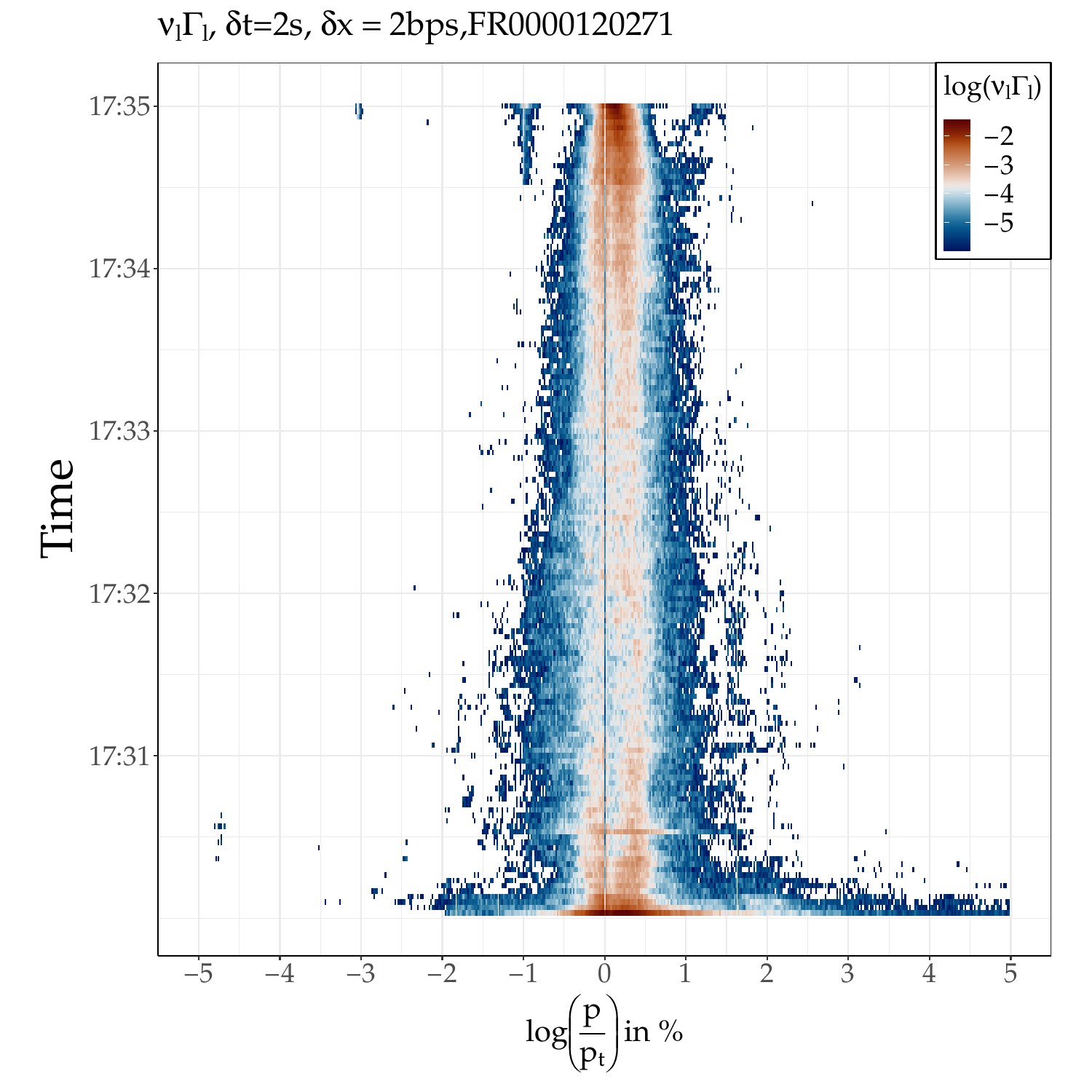}
    \includegraphics[scale=0.295]{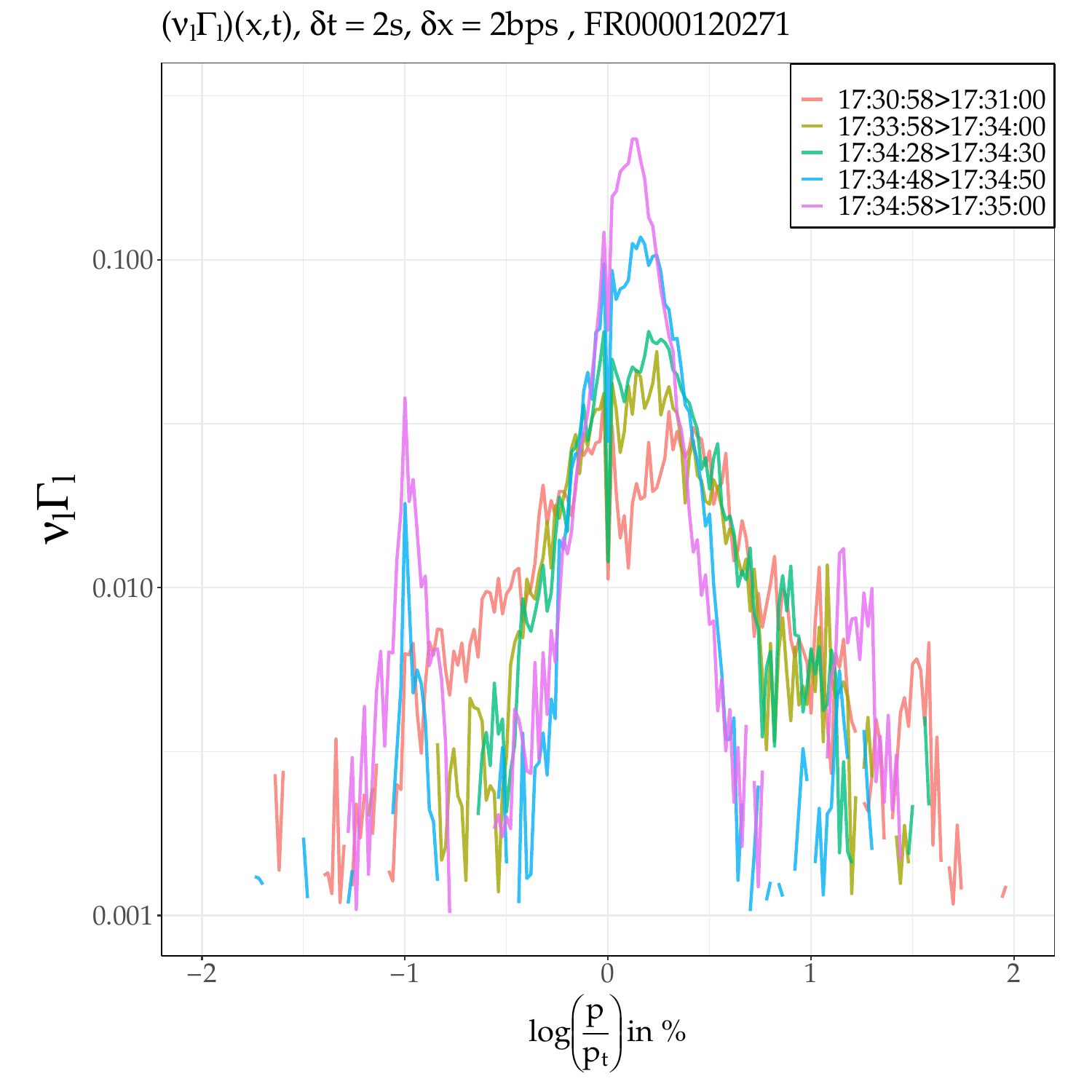}
    \caption{Estimating the cancellation rate $\nu_l\Gamma_l$ from tick by tick data. Left panel: heatmap  $(x,t) \rightarrow \nu_l\Gamma_l$. Right panel: $x \rightarrow\nu_l\Gamma_l$ at $n$-seconds before auction time $T=$17:35:00, $n \in \{0,10,30,60,240\}$.}
    \label{fig:Gamma_L_1}
\end{figure}

\begin{figure}
    \centering    
    \includegraphics[scale=0.295]{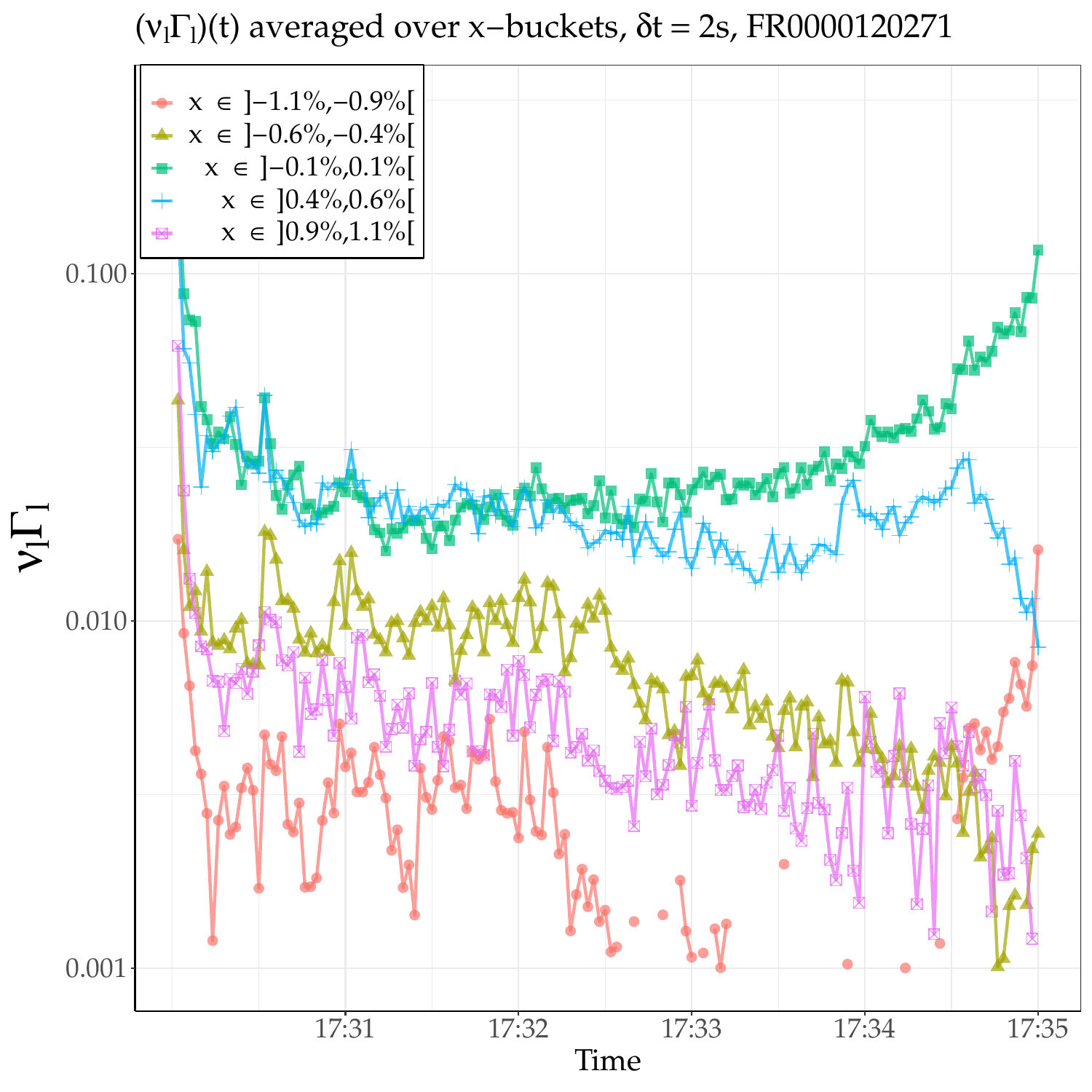}
    \includegraphics[scale=0.295]{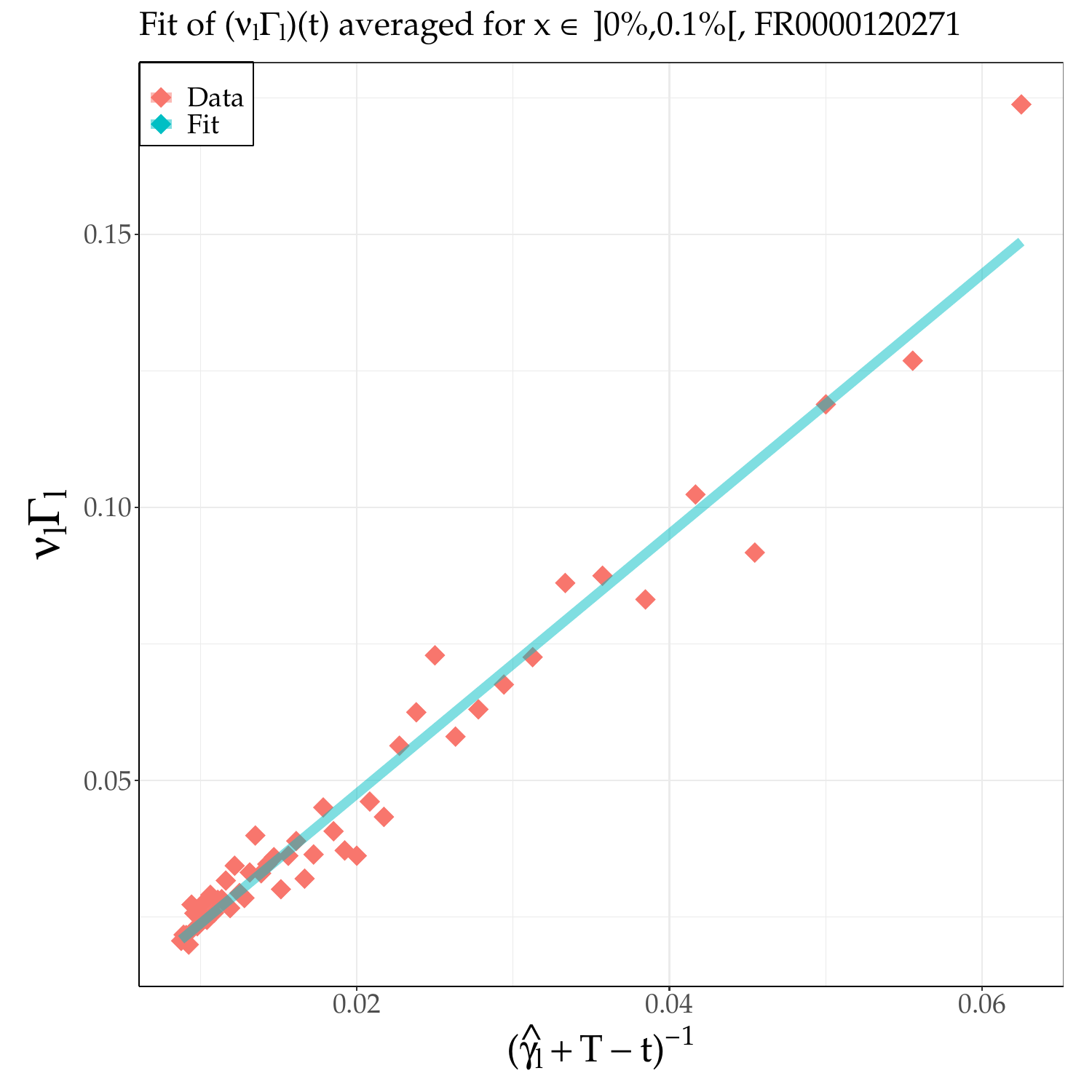}
    \caption{Estimating $\nu_l\Gamma_l$. Left panel: $t \rightarrow \left<\nu_l\Gamma_l\right>|_{x}$ averaged for different price buckets as a function of time. Right panel: average values of $\nu_l\Gamma_l$ for x in $]0,0.1\%[$ (diamonds) as a function of $(\gamma_l + T-t)^{-1}$ for $t>t_l^{(0)}$. The solid line is a linear fit minimizing the L1 norm of errors. Fitted values: $t_l^{(0)} = 202$, $\hat{\gamma_l} = 16$,  $\hat{C_l} = 2.38$.}
    \label{fig:Gamma_L_2}
\end{figure}

For each aggregation period $[t-\delta t, t]$ and each asset, we average Eq. \eqref{eq:estim_gamma_l} over all the days in our dataset using equal log price intervals of length $\delta x = 2$ basis points and time step $\delta t =2$ seconds. Figure \ref{fig:Gamma_L_1} shows a non-trivial behavior of the cancellation rate w.r.t. $x$ and $t$. At the beginning of the auction, a series of order cancellations yields a decreasing cancellation rate as a function of time. In addition, we have $\nu_l\Gamma_l \propto e^{-|x|}$. As the auction time approaches, $\nu_l\Gamma_l$ reaches a maximum at a value larger than the indicative price.
We report in the left panel of Fig. \ref{fig:Gamma_L_2} averaged values of $\nu_l\Gamma_l$ as a function of time over different price buckets in order to display the temporal dependency of cancellations: we confirm the cancellation decrease at the start of the auction, then the increase for prices around indicative price. We verify that $\nu_l\Gamma_l\propto 1/(\gamma_l + T-t)$ for $0<x<10$bps and  $t>t_l^{(0)}$. To determine  $t_l^{(0)}$, we use a change point detection criterion: we assume that when $t<t_l^{(0)}$ a constant model is a better fit than a fit to $t \rightarrow 1/(\gamma_l + T-t)$ and vice versa when $t>t_l^{(0)}$. We exclude the first minute of the auction when cancellations decrease. The right panel of Fig. \ref{fig:Gamma_L_2} displays such a time fit for TotalEnergies.

\begin{figure}
    \centering
    \includegraphics[scale=0.295]{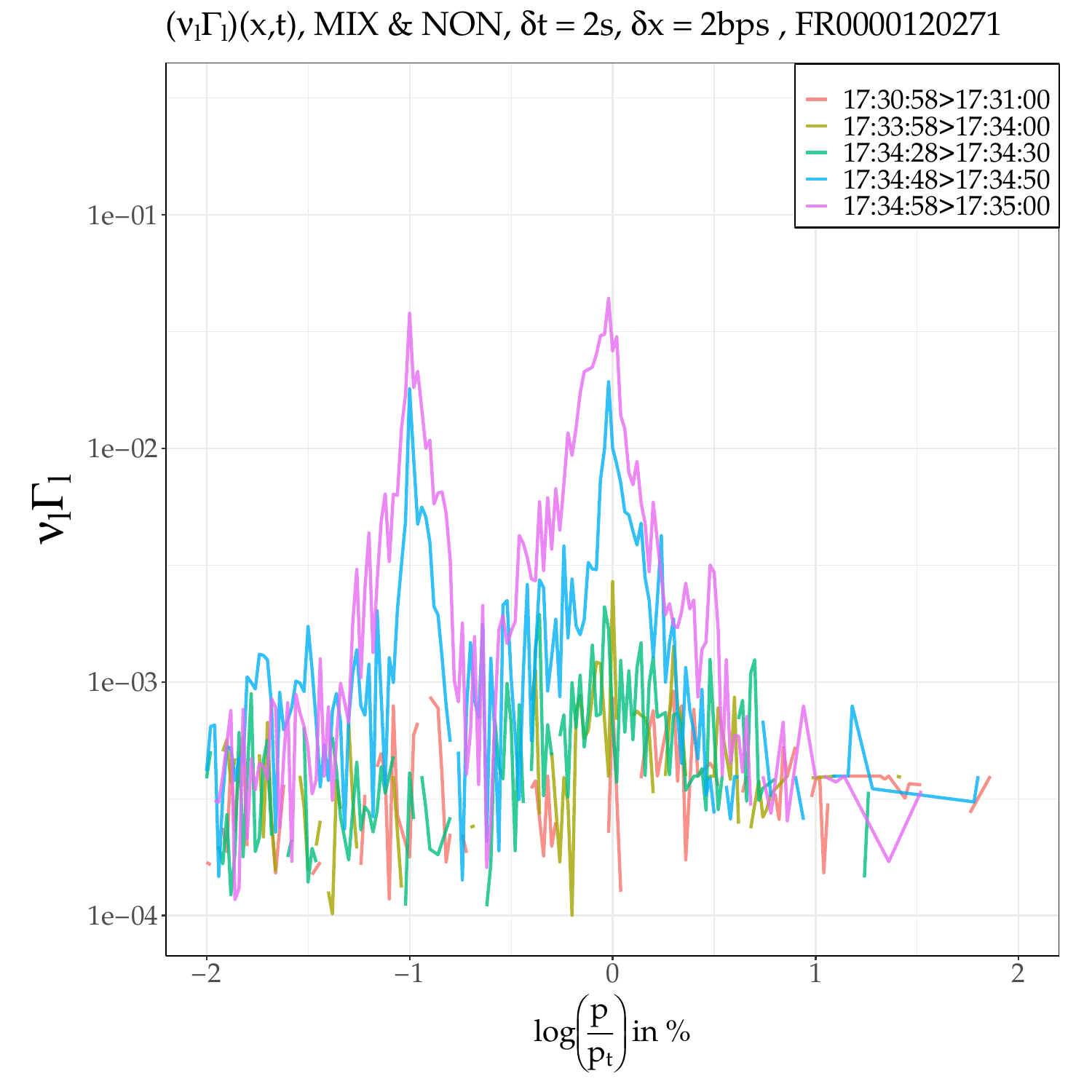}
    \includegraphics[scale=0.295]{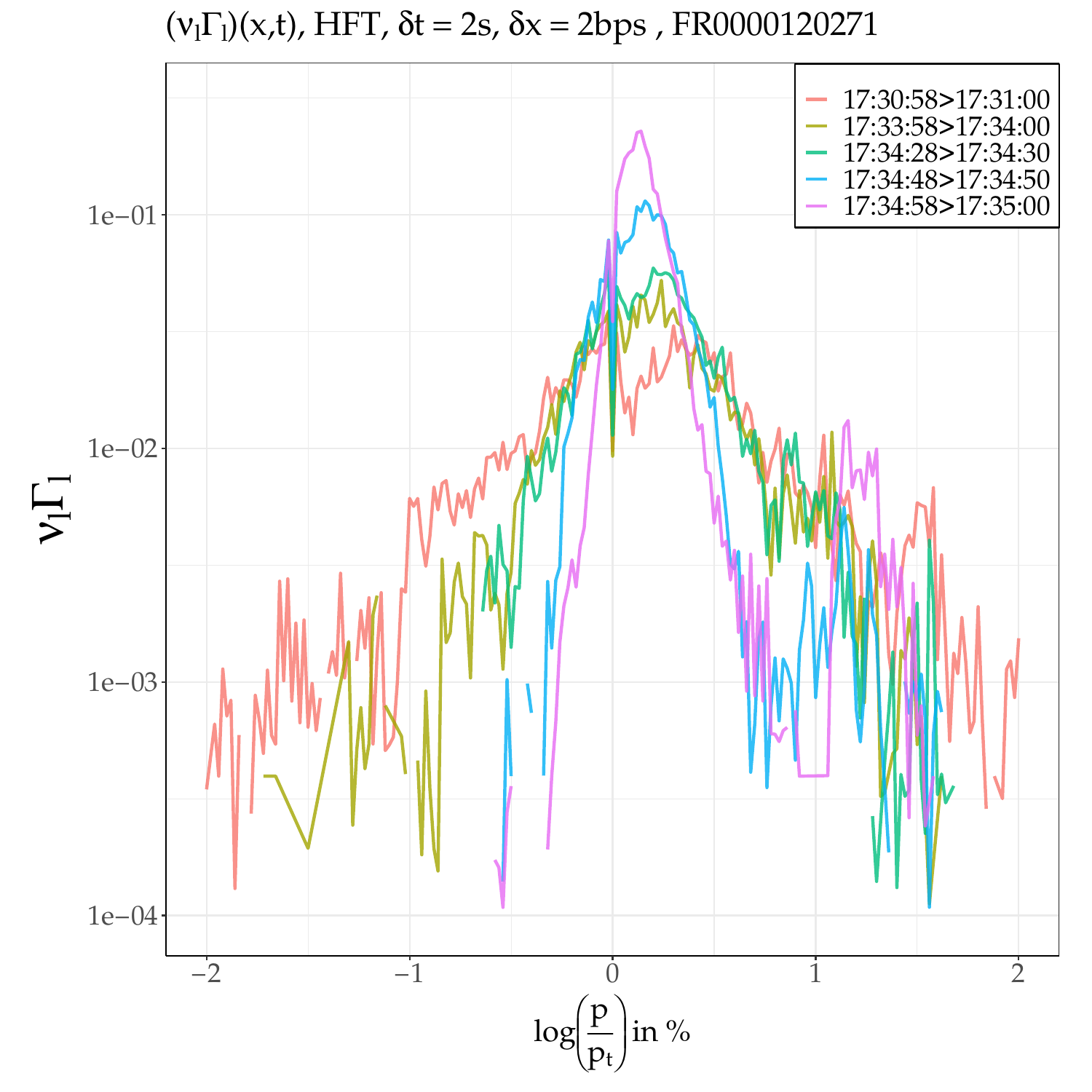}
    \caption{Estimating the cancellation rate across market participants. Left panel: MIX \& NON contribution. Right panel: HFT contribution.}
    \label{fig:Gamma_L_3}
\end{figure}

To examine the variation in the cancellation rates across market participants, we compute the cancellation rate separately for HFT-flagged traders and for non-HFTs. Although MIX includes the high-frequency activity of investment banks, we denote MIX and NON as non-HFTs for simplicity. Figure \ref{fig:Gamma_L_3} reveals that the shape of cancellations for non-HFTs is markedly different from that of HFTs. The cancellation rate of non-HFTs is noisier and is only weakly dependent on the price most of the time, then peaking around the indicative price with exponential decay as the auction time approaches. On the contrary, HFT-flagged traders are the predominant contributors to the cancellation rate, significantly outweighing the cancellations initiated by non-HFTs. In reality, HFTs are highly active in auctions, accounting for more than 80\% of all events on average, yet they contribute to only a minor portion of the closing volume, approximately 4\% on average.
 
Lastly, notice a significant peak at $x=-1\%$ indicating a strategic behavior of non-HFTs that are canceling, on average, large matched volumes a few seconds prior to auction time.

\subsubsection{Measuring the submission rate: $\nu_r\Gamma_r$}
\label{sec:mes_gamma_r}
The product of the submission rate $\nu_r\Gamma_r$ by the latent order density $\rho^{(l)}$ for price $p$ between $t-\delta t$ and $t$ is defined as 
\begin{equation}
    (\nu_r\Gamma_r \cdot \rho^{(l)})(p, t -\delta t \rightarrow t) = \frac{1}{\delta t}\sum_{t-\delta t<t_i<t}
    \frac{\delta V(p,t_i)}{\delta x \cdot Q_a}\cdot \mathbb{1}_{\{\delta V(p,t_i)>0\}},
\label{eq:estim_gamma_r}
\end{equation}
where $\delta V(p,t_i)$ is the volume change in limit price $p$ at time $t_i$ excluding volumes that diffused from another price $p'$, and $Q_a$ is the final auction volume:  we scale the submitted density $\delta V(p,t_i)/\delta x$ by the final auction volume $Q_a$ to be able to average Eq.\ \eqref{eq:estim_gamma_r} over different days.

\begin{figure}
    \centering
    \includegraphics[scale=0.295]{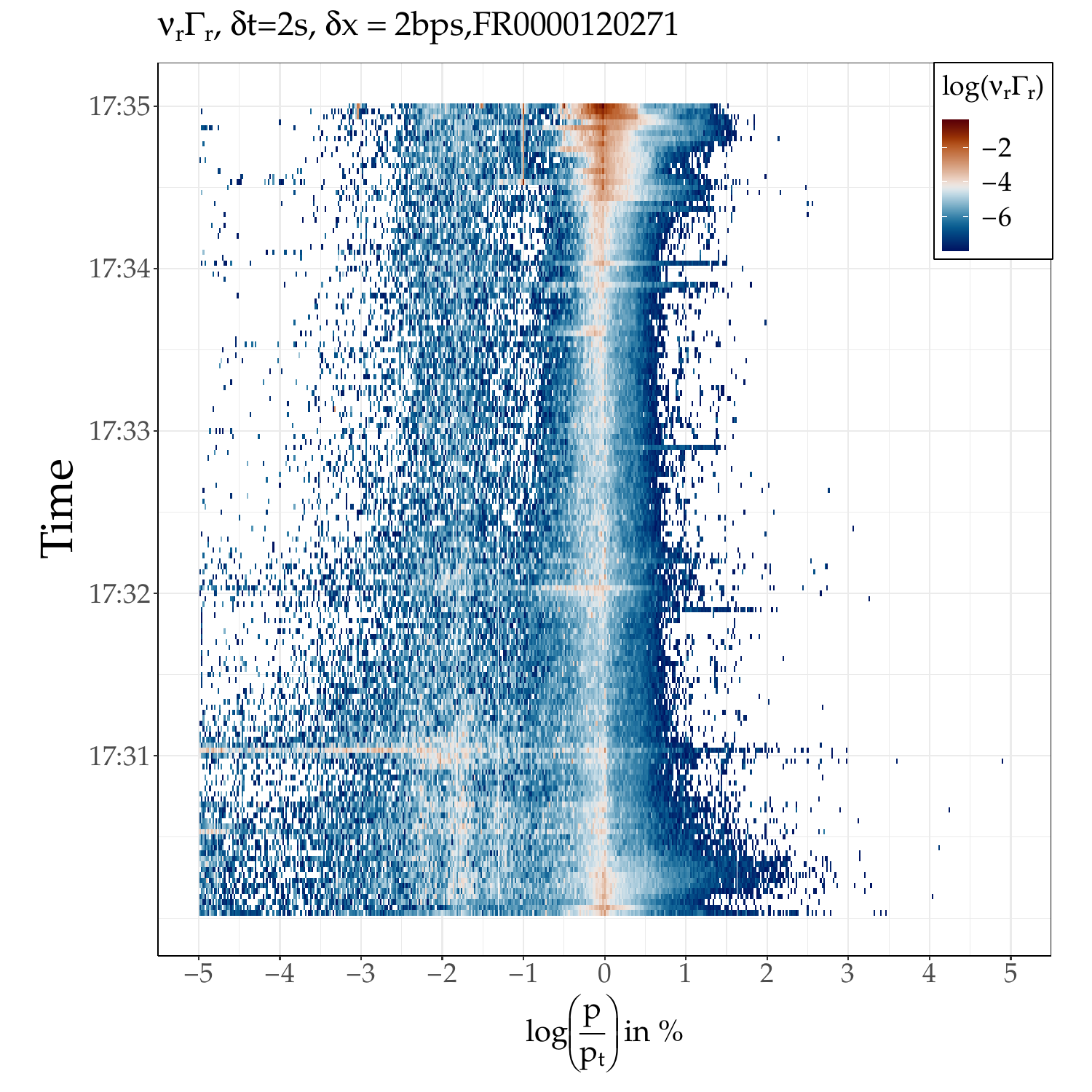}
    \includegraphics[scale=0.295]{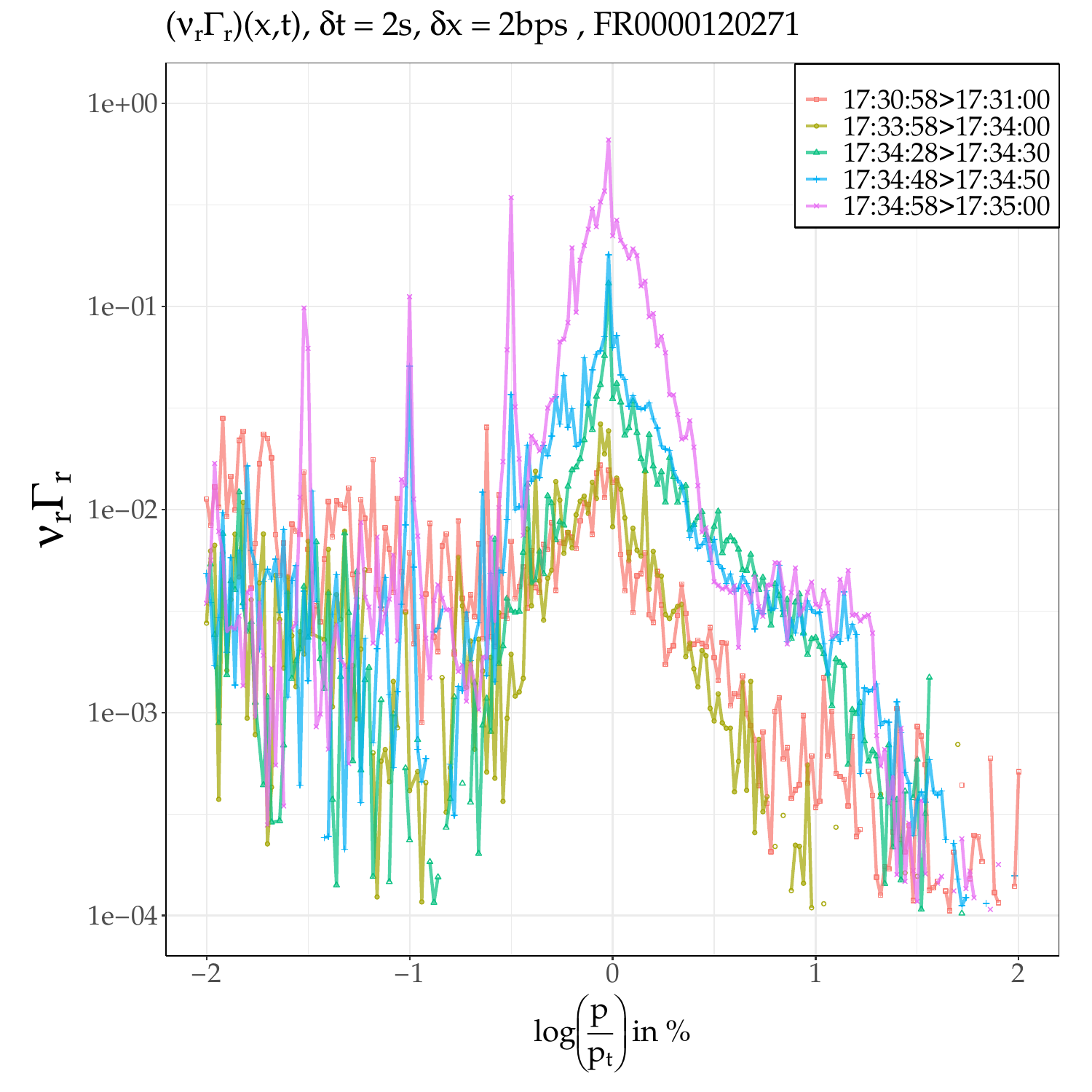}
    \caption{Estimating the submission rate $\nu_r\Gamma_r$ from tick by tick data. Left panel: heatmap $(x,t) \rightarrow\nu_r\Gamma_r$. Right panel: $x \rightarrow\nu_r\Gamma_r$ at $n$-seconds before auction time $T=$17:35:00, $n \in \{0,10,30,60,240\}$.}
    \label{fig:Gamma_R_1}
\end{figure}

Likewise, we average Eq. \eqref{eq:estim_gamma_r} over days with equal log price intervals of $\delta x = 2$ basis points and time step $\delta t = 2$ seconds. Then, we infer $\nu_r\Gamma_r$ using the numerical estimate of the latent order book from section \ref{sec:static_fits} where $\rho^{(l)}(x) = \max(\hat{a},\hat{a}x+\hat{b})$. We implicitly assume that the latent order book remains stable throughout the accumulation period. This is the case when the latent book is significantly larger in comparison with the revealed order book at all times. Figure \ref{fig:Gamma_R_1} shows that $\nu_r\Gamma_r$ displays a clear exponential decay for $x>0$ with a constant price scale at most times, and a truncated exponential decay for $x<0$.

\begin{figure}
    \centering
    \includegraphics[scale=0.29]{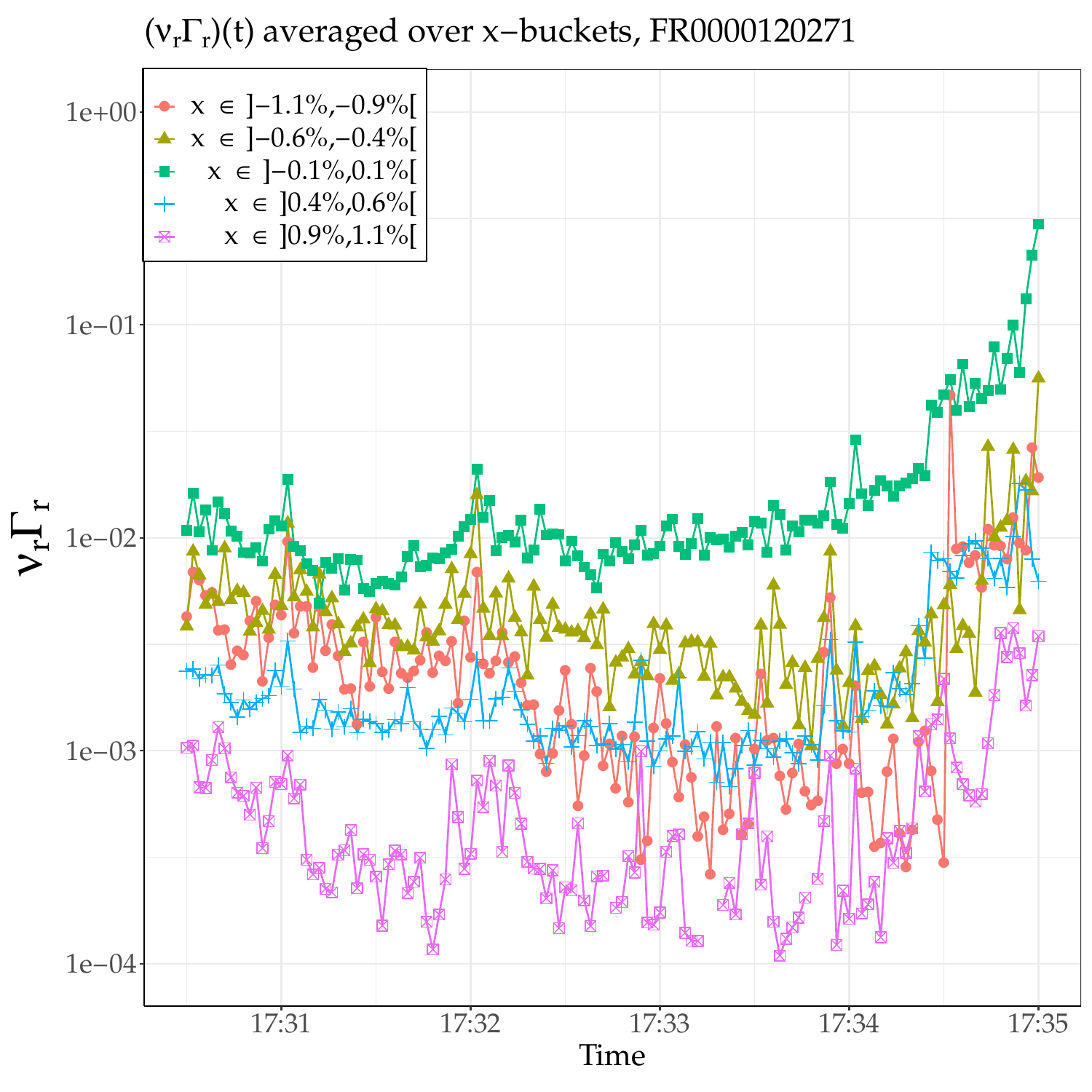}
    \includegraphics[scale=0.29]{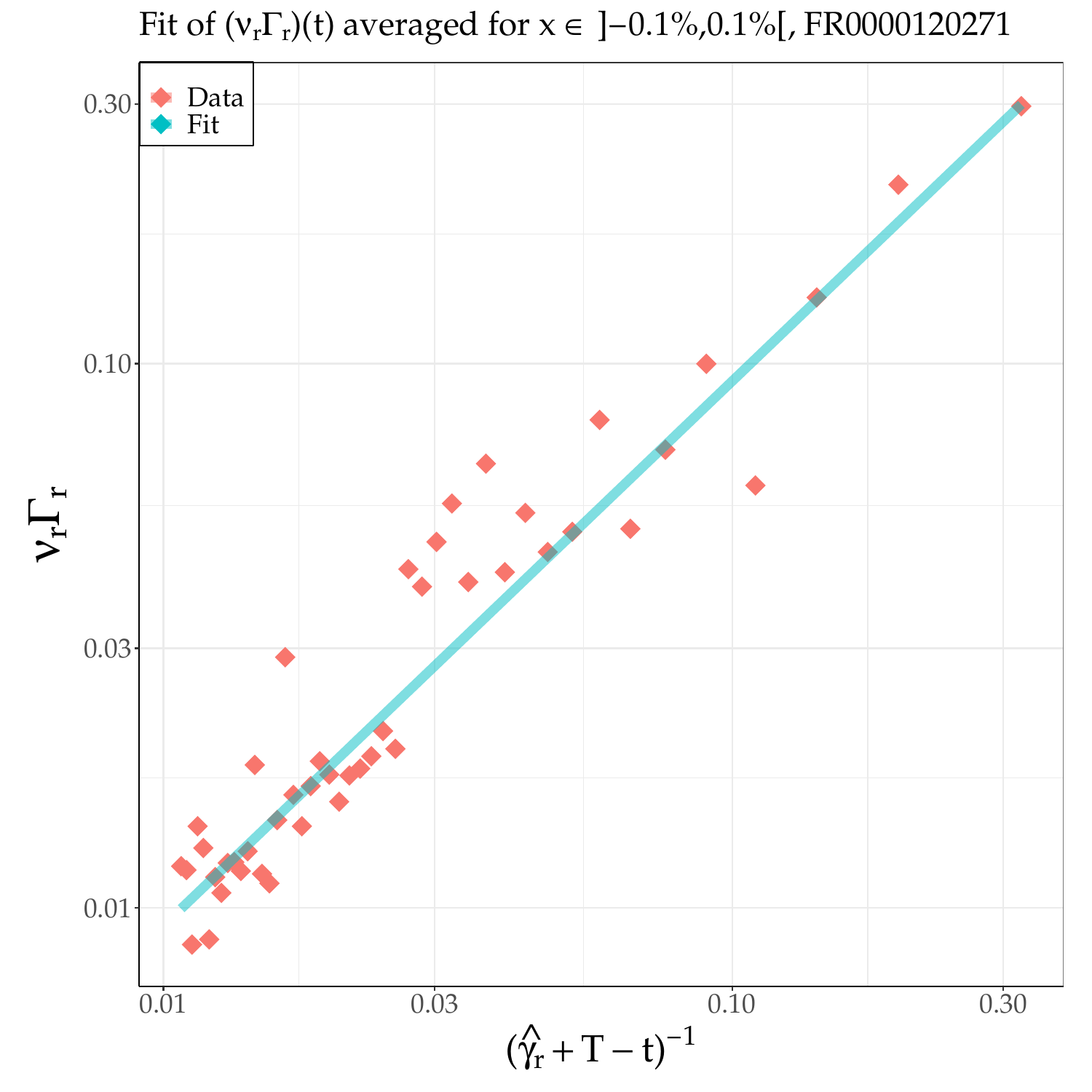}
    
    \caption{Estimating $\nu_r\Gamma_r$. Left panel: $t \rightarrow \left<\nu_r\Gamma_r\right>|_{x}$ averaged for different price buckets as a function of time. Right panel: average values of $\nu_r\Gamma_r$ for x in $]-0.1,0.1\%[$ (diamonds) as a function of $(\gamma_r + T-t)^{-1}$ for $t>t_r^{(0)}$. The solid line is a linear fit minimizing the L1 norm of errors. Fitted values: $t_r^{(0)} = 210$, $\hat{\gamma_r} = 3.1$,  $\hat{C_r} = 0.93$.}
    \label{fig:Gamma_R_2}
\end{figure}

We report in the left panel Fig. \ref{fig:Gamma_R_2} averaged values of $\nu_r\Gamma_r$ as a function of time over different price buckets to display the temporal dependency of submissions. We empirically verify that $\nu_r\Gamma_r$ grows proportionally to $1/(\gamma_r + T-t)$ for values of $t>t_r^{(0)}$, and determine $t_r^{(0)}$ using a change detection criterion. We report in the right panel of Fig. \ref{fig:Gamma_R_2} such a time fit for TotalEnergies.

\begin{figure}
    \centering
    \includegraphics[scale=0.295]{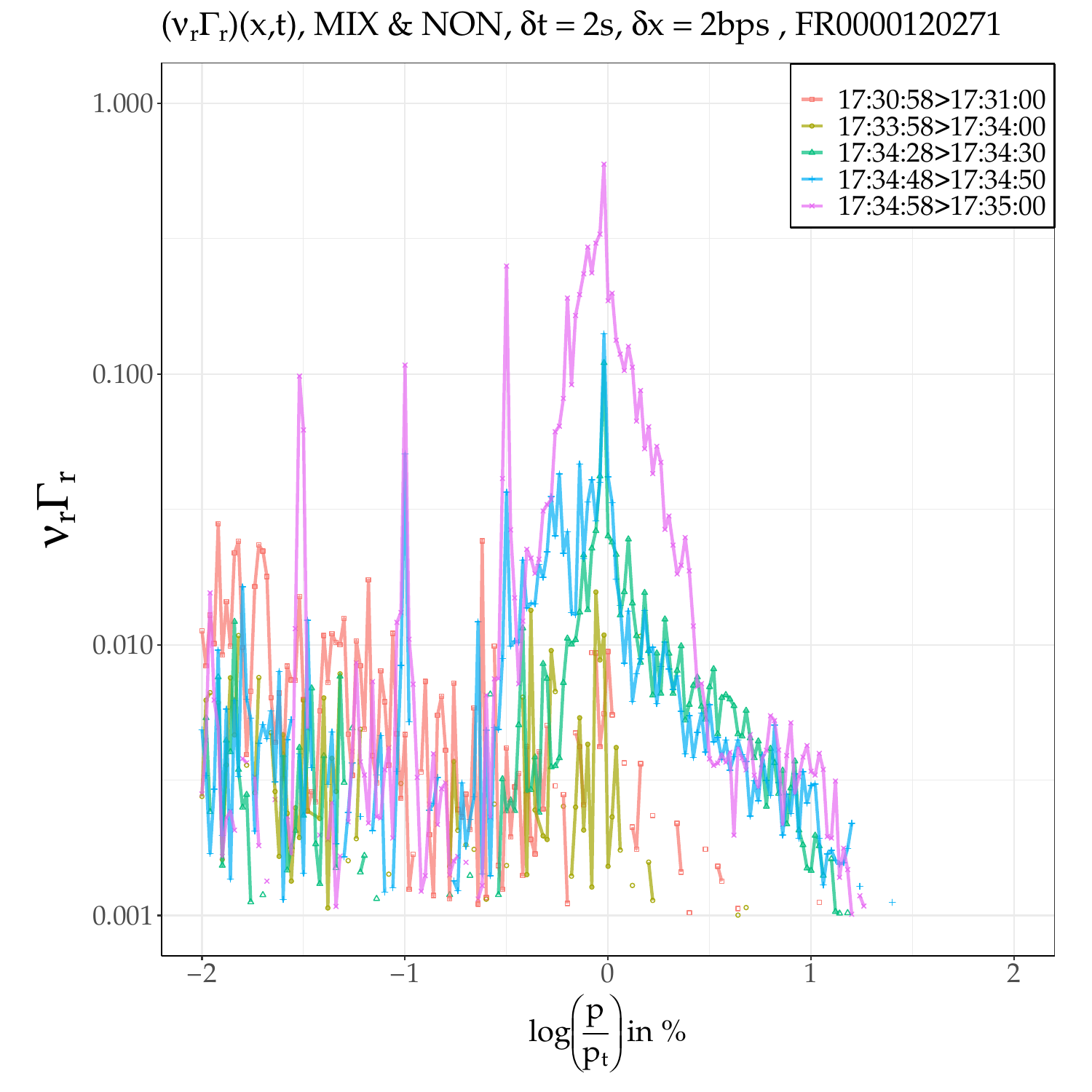}
    \includegraphics[scale=0.295]{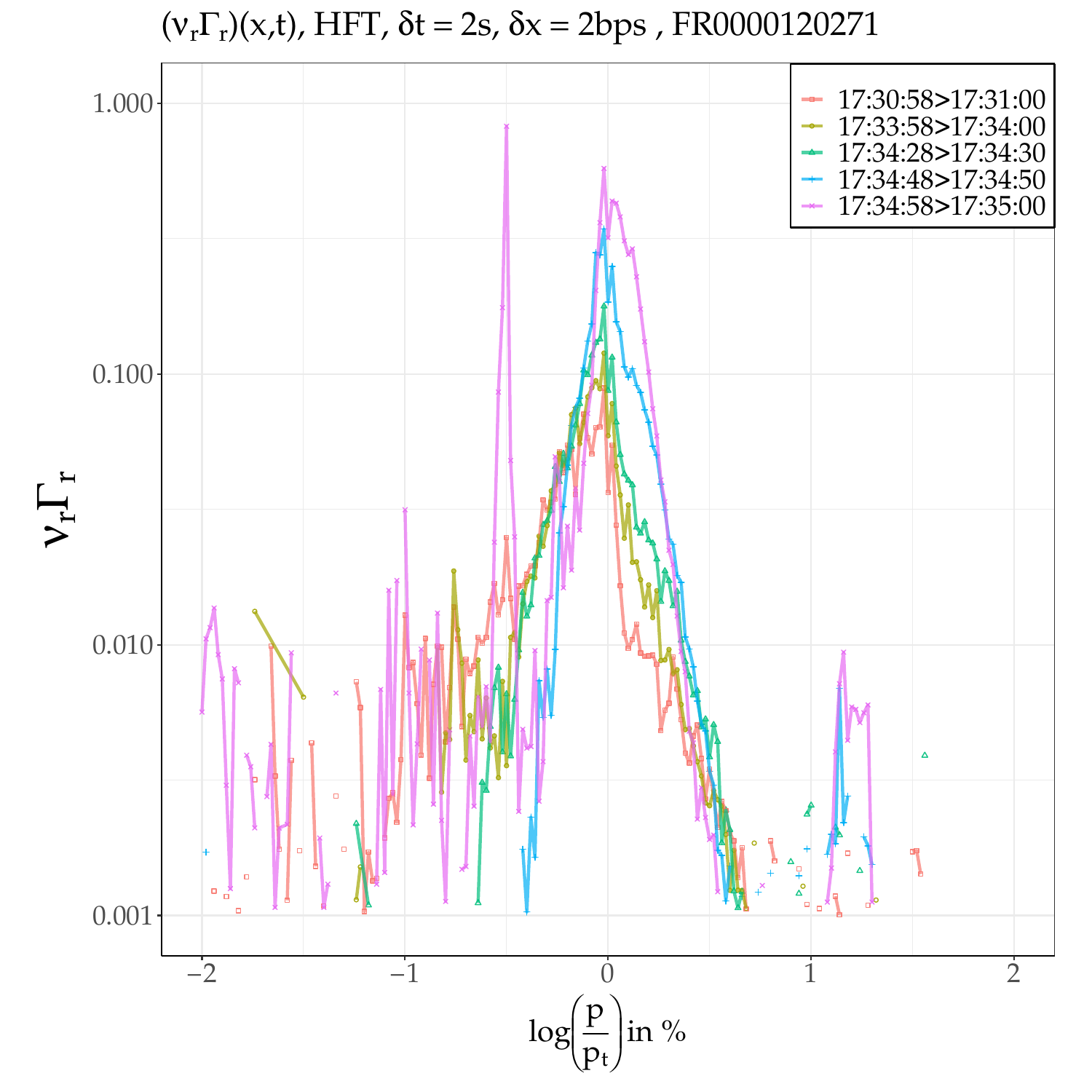}
    \caption{Estimating the submission rate across market participants. Left panel: MIX \& NON contribution. Right panel: HFT contribution.}
    \label{fig:Gamma_R_3}
\end{figure}

Considering HFTs and non-HFTs separately in Fig. \ref{fig:Gamma_R_3}, we find a similar shape of submissions for both with a larger price scale for non-HFTs. Note that we inferred each submission rate (HFTs and non-HFTs) from the respective latent book as computed section \ref{sec:static_fits}. Lastly, we notice large peaks at $x=0$ and at multiples of $-0.5\%$ as the clearing approaches: these point to agents trying to pin the current indicative price when they send orders at $x=0$, or to construct a barrier of matchable orders at a less favorable price when they send them at $x<0$.

\subsubsection{Measuring diffusion in the revealed order book: $D_r$}
\label{sec:mes_diffusion} 
When prices are diffusive, the diffusion coefficient \cite{donier2016walras} $D(x)$ at a price $x \in \mathbb{R}$ is defined as
\begin{equation}
    D(x) = \frac{1}{2}\left(\text{Var}(\beta)\cdot \sigma^2   + \int_{\mathbb{R}} (x-y)^2\Gamma_D(x,y)\mathrm{d}y\right),
    \label{eq:diff_coef}
\end{equation}
where $\sigma$ is the volatility of the indicative price, $\text{Var}(\beta)$ is a prefactor that encodes the heterogeneity of agent reactions to price movements: should agents neither over-react nor under-react to indicative price movements ($\beta = 1$ for all agents), the first term vanishes;
\citet{toth2011anomalous} consider a unit prefactor $\text{Var}(\beta)= 1$. The other term  $\int (x-y)^2 \Gamma_D(x, y)dy$ corresponds to $\mathbb{E}_i[\Sigma_i^2]$ in \citet{donier2016walras} and represents ``the purely idiosyncratic noisy updates of agents'': here $\Gamma_D(x,y)$ is the rate at which orders are updated from log-price $x$ to log-price $y$. Historically, the derivation of the diffusion coefficient in latent models was introduced first in \citet{toth2011anomalous}, then developed further in the appendix of \citet{donier2015fully}. 

The definition of Eq. \eqref{eq:diff_coef} corresponds to that of the revealed diffusion coefficient $D_r$ in our framework, provided we only account for visible price reassessments. Note that the under-diffusive nature of the indicative price may result in a different first-term contribution to the diffusion coefficient. 

\begin{figure}
    \centering
    \includegraphics[scale=0.19]{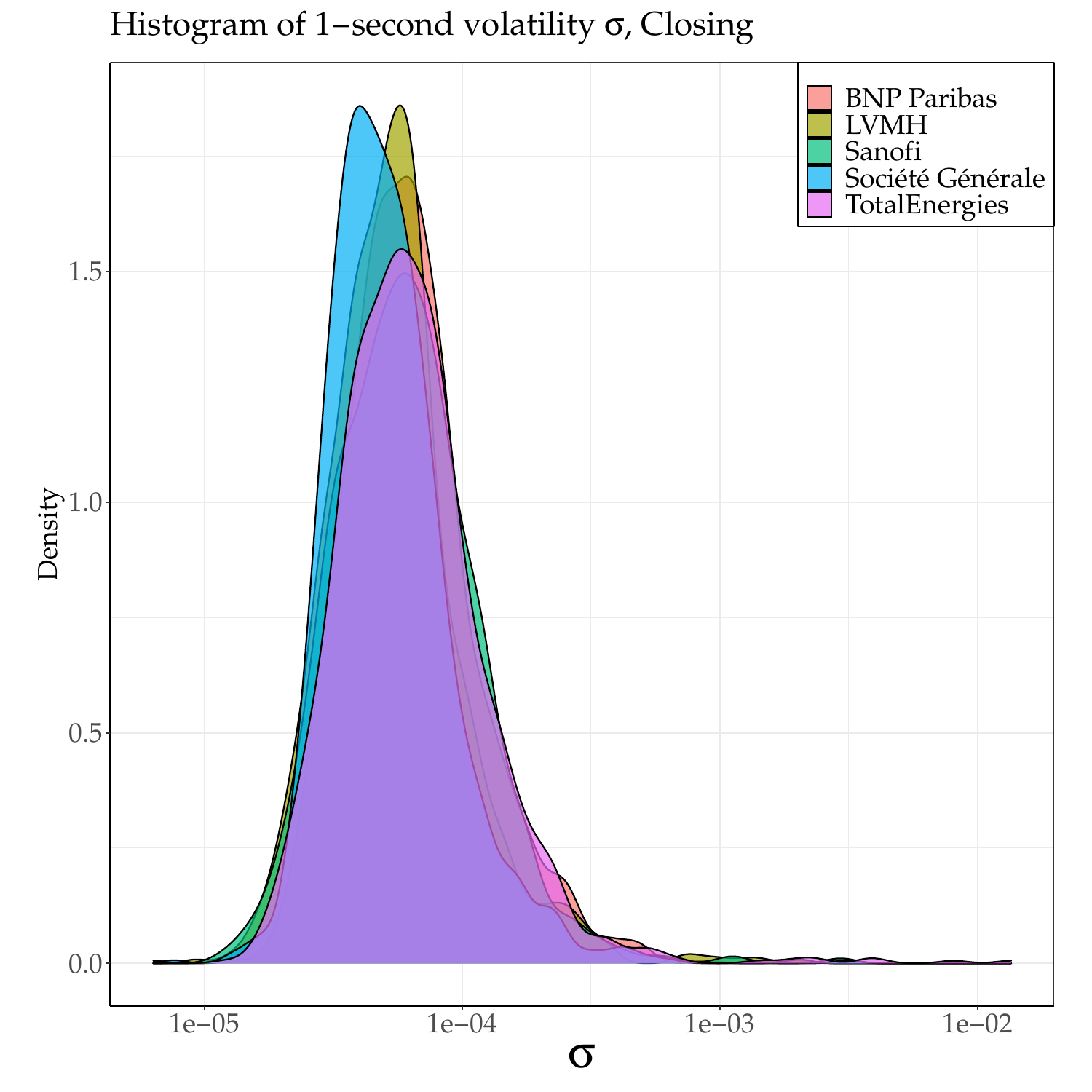}
    \includegraphics[scale=0.2]{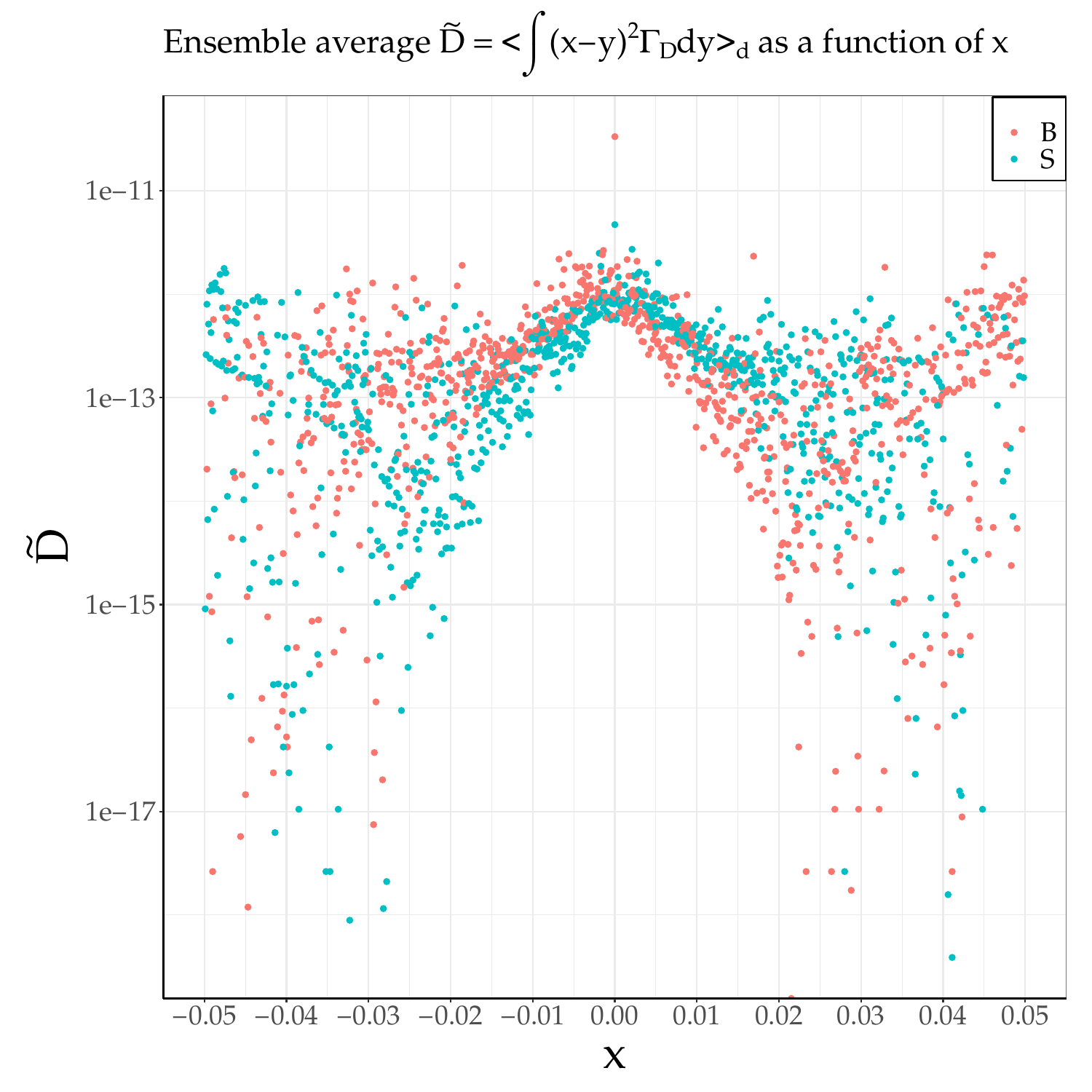}
    \includegraphics[scale=0.195]{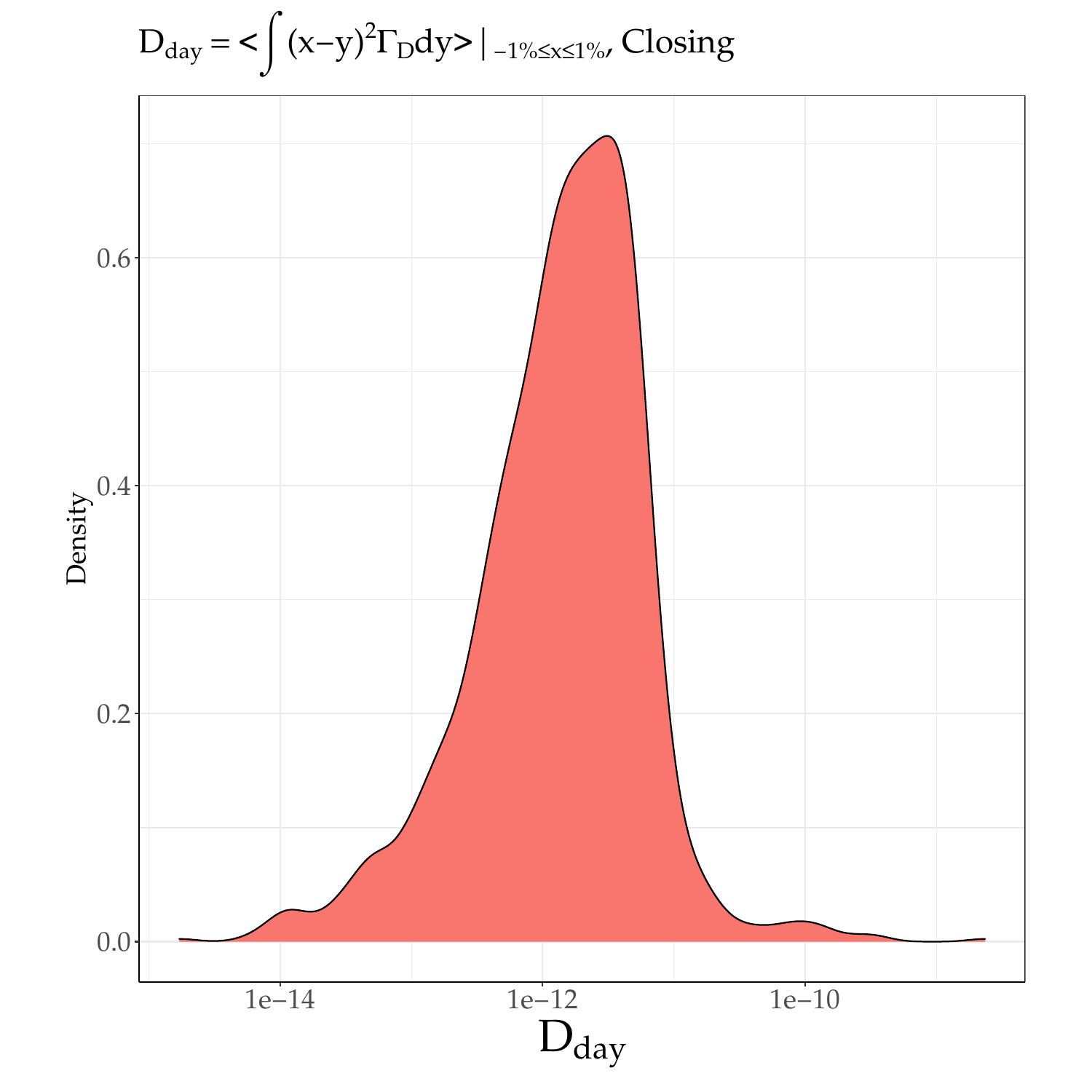}
    \caption{Left panel: histogram of daily 1-second volatility during the closing auction. Middle panel: ensemble average of the price update contribution to diffusion as a function of $x$. Right panel: histogram of daily price update contribution (restricted to limit prices $-1\% \leq x \leq 1\%$).}
    \label{fig:diff_measures}
\end{figure}

First, let us consider the volatility per unit of time $\sigma$. The average 1-second volatility is of order $10^{-3}$ in the first minute of the accumulation period and of order $10^{-4}$ in the last four minutes (see section \ref{sec:anomalous_scaling}).
To inspect daily measures, we plot in the left panel of Fig. \ref{fig:diff_measures} a histogram of the daily realized 1-second volatility $\sigma = \sqrt{(\sum_{t=1}^N \log(p_{t+1\text{s}}/p_t)^2)/N}$ during an auction, with medians of order $10^{-4}$. On an average day, the volatility contribution in the diffusion coefficient is of order $10^{-8}$.

Let us now measure the contribution of pure price updates $\frac{1}{2}\int_{\mathbb{R}} (x-y)^2\Gamma_D(x,y)\mathrm{d}y$. We compute the update rate $\Gamma_D(x,y)$ as
\begin{equation}
    (\Gamma_D)(x, y, t -\delta t \rightarrow t)= \frac{1}{\delta t}\sum_{t-\delta t<t_i<t}
    \frac{\delta V(p, p', t_i)}{V_p(t_i)},
\end{equation}
where $\delta V(p, p', t_i)$ is the amount of volume moving from limit price $p$ to $p'$ at time $t_i$, $x = \log(p/p_{t_i}^-)$, $y = \log(p'/p_{t_i}^+)$, and $V_p(t_i)$ is total volume in limit price $p$ at time $t_i^-$.
We plot in Fig. \ref{fig:diff_measures} the average contribution (across days) of the second term $\left<\frac{1}{2}\int_{\mathbb{R}} (x-y)^2\Gamma_D(x,y)\mathrm{d}y\right>_d(x)$ as a function of $x$ (middle panel), and a histogram of daily values (for $-1\% \leq x \leq 1\%$ ) of average price reassessments $\left<\frac{1}{2}\int_{\mathbb{R}} (x-y)^2\Gamma_D(x,y)\mathrm{d}y\right>|_{-1\% \leq x \leq 1\%}$ (right panel). These measurements show that the second term has a mode of order $ 10^{-12}$.
Thus, the contribution of price reassessments in the revealed diffusion coefficient is negligible in comparison with that of the indicative price volatility provided unit prefactor, in line with the findings of \citet{challet2001analyzing}.

It is worth noting that visible price reassessments might be undervalued, as agents may choose to cancel a limit order entirely before resubmitting it at an updated price. This mechanism, which is difficult to observe in our data, should be accounted for as a form of revealed diffusion. Time priority is less important in auctions than in continuous trading; it is important only for limit orders whose price at the auction time equals the auction price.

\subsection{Solving the full dynamical equations}
\label{sec:num_solving}

To replicate the full dynamics of the revealed auction book throughout the accumulation period, we return to the full model introduced in section \ref{sec:auction_model} and solve Eqs \eqref{eq:modelS} numerically. To this end, we take a constant cancellation rate $(\nu_l\Gamma_l)(x,t) = \nu_l>0$, and model the submission rate $\nu_r\Gamma_r$ as follows: For positive prices $x \geq 0 $, we take a weighted sum of two exponential terms, representing, respectively, the contribution of fast agents with price scale $x_r$ and slow agents with price scale $k\cdot x_r$, ($k>1$) (see section \ref{sec:static_fits}). We allow for the fast agent contribution to increase proportionally to $1/(\gamma_r +T -t)$ when $t>t_r^{(0)}$. For negative prices, we choose a constant submission rate for prices $x<-x_0$, where $x_0>0$ is a threshold, then we employ one exponential term for prices $-x_0 \leq x <0 $ establishing continuity at $x=0$ and $x=-x_0$. This yields
\begin{equation}
    (\nu_r\Gamma_r)(x,t) = 
    \begin{cases}
         \frac{w\cdot C_r}{\gamma_r+T-\max\{t,t_r^{(0)}\}}\cdot e^{-x/x_r} + \frac{ (1-w)\cdot C_r}{\gamma_r+T-t_r^{(0)}}\cdot e^{-x/(k\cdot x_r)} \quad &,x\geq 0; \\
         A^*\cdot e^{x/x_r^*} &, -x_0 \leq x <0;\\
         m \cdot\frac{ C_r}{\gamma_r} & ,x<-x_0,
    \end{cases}
    \label{eq:nu_r_param}
\end{equation}
where $C_r, x_r, x_0, m>0$, $0<w<1$, and $\gamma_r,t_r^{(0)}$ are computed in section \ref{sec:mes_gamma_r}. $A^*$ and $x_r^*$ are determined by the continuity conditions at $x=0$ and $x=-x_0$. We complement Eqs \eqref{eq:modelS} with the following initial conditions
\begin{equation}
    \begin{aligned}
        \rho^{(r)}(x,0) &= 0;\\
        \rho^{(l)}(x,0) &= \max(ax+b,b),
    \end{aligned}
\end{equation}
where $a,b>0$ are the stationary latent book parameters computed in section \ref{sec:static_fits}. 

To calibrate Eqs \eqref{eq:modelS} to order book data, we minimize the sum of squared errors at times 0,10,..., 290 seconds before the auction time $t \in J = \{10, 20, \dots, 300 \}$ and prices $x \in I = [-2\%,2\%]$.
\begin{equation}
    f(C_r,x_r,k,w,\nu_l, x_0, m, \dots ) = \sum_{ x \in I}\sum_{  t \in J } \left(\rho(x,t)-\widehat{\rho(x,t)}\right)^2.
    \label{eq:obj_func}
\end{equation}

Finally, to measure the influence of diffusion, we minimize \eqref{eq:obj_func} first taking $D_r = D_l = 0$ (Zero diffusion), then we allow for $D_r,D_l>0$ as constant parameters to be calibrated (Constant diffusion). Lastly, as suggested by the temporal pattern of the indicative price volatility of section \ref{sec:anomalous_scaling}, we opt for a time-dependent $D_r$ (Time diffusion) 
\begin{equation}
    D_r(t) =
    \begin{cases}
         D_0&, \quad 0< t \leq 1;\\
         (D_T-D_0)\cdot\frac{t^{-1}-1}{T_s^{-1}-1} + D_0&, \quad 1 <  t< T_s  ;\\   
         D_T&,\quad  t \geq T_s.
    \end{cases}
    \label{eq:D_r_param}
\end{equation}
Eq. \eqref{eq:D_r_param} states that the revealed diffusion coefficient is equal to $D_0$ at times $ 0< t \leq 1$ then decreases as a power law with exponent -1 to a value $D_T<D_0$ until time $t=T_s$, where $T_s\leq T$ is a saturation time. Then for times $t\geq T_s$, $D_r$ remains equal to $D_T$. We take $T_s = 180$s.

\begin{figure}
    \centering
    \includegraphics[scale=0.5]{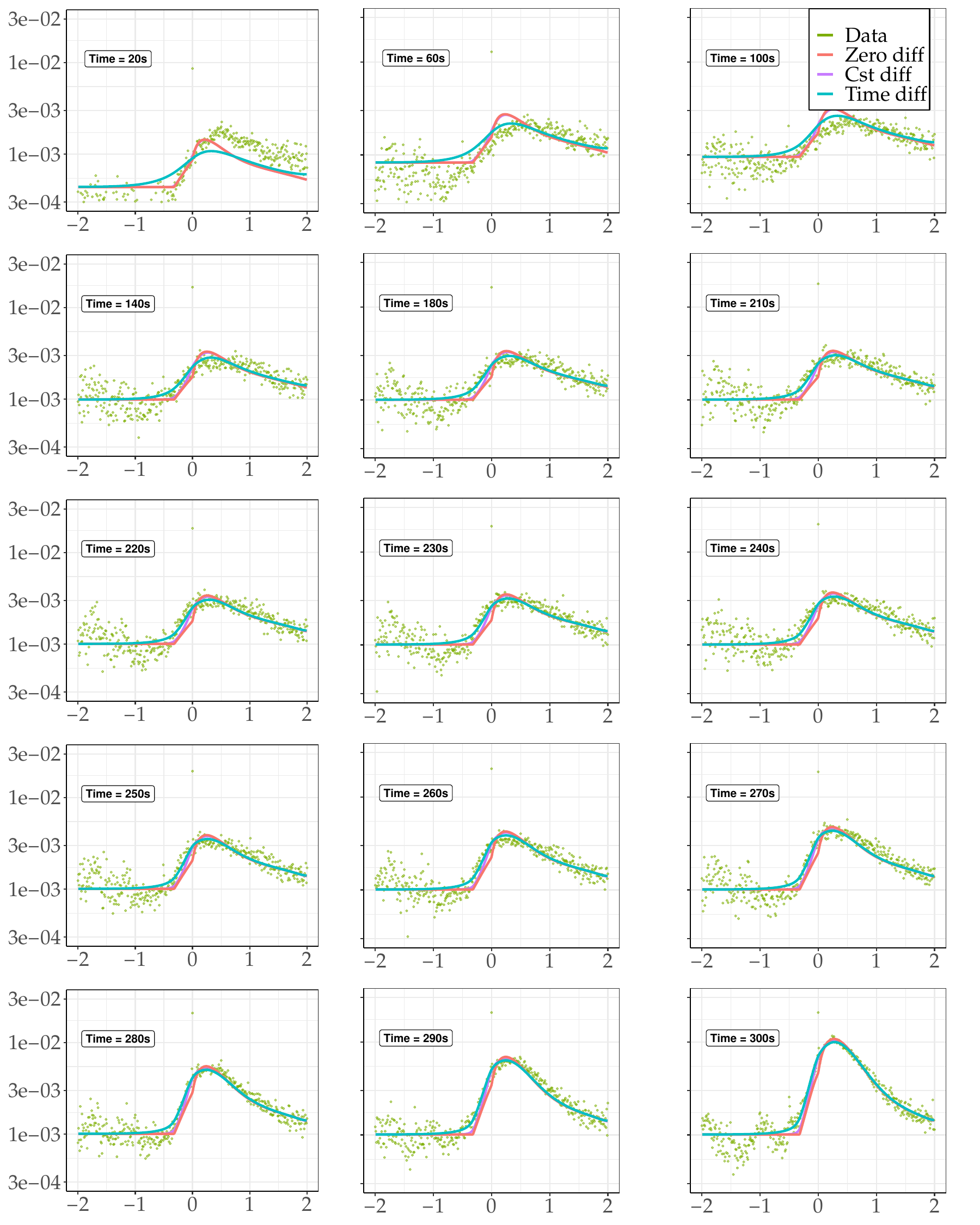}
    \caption{Numerical fits $\rho^{(r)}_S$ (solid lines) of the average auction book (green dots) as a function of the centered log price $-2\% \leq x \leq 2\%$ at different instants during the accumulation period. The Y-axis is in log scale. The zero diffusion fit ($D_r = D_l =0$) is in red lines, the constant diffusion fit ($D_r,D_l>0$) is in purple lines, and the time diffusion fit ($D_r \propto 1/t$) is in blue lines.}
    \label{fig:num_sol_as_function_of_price}
\end{figure}

\begin{figure}
    \centering
    \includegraphics[scale=0.5]{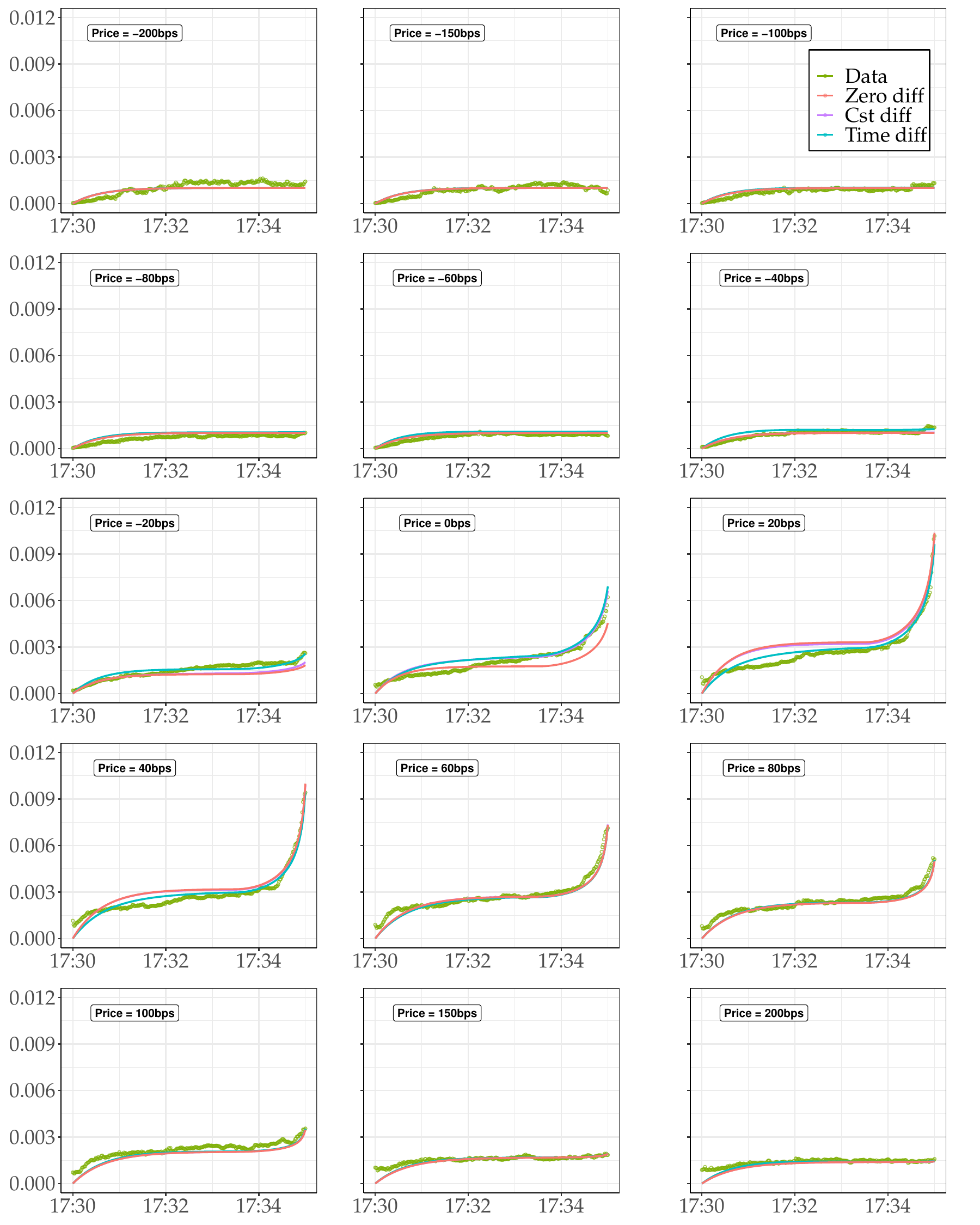}
    \caption{Numerical fits $\rho^{(r)}_S$ (solid lines) of the average auction book (green dots) as a function of time $0<t\leq T$ at different prices price $-2\% \leq x \leq 2\%$. The Y-axis is in ordinary scale. The zero diffusion fit ($D_r = D_l =0$) is in red lines, the constant diffusion fit ($D_r,D_l>0$) is in purple lines, and the time diffusion fit ($D_r \propto 1/t$) is in blue lines.}
    \label{fig:num_sol_as_function_of_time}
\end{figure}

We present the obtained fits first in the plane $(\rho^{(r)},x)$ at different times in Fig. \ref{fig:num_sol_as_function_of_price}, then in the plane $(\rho^{(r)},t)$ at different prices in Fig. \ref{fig:num_sol_as_function_of_time}. Even under a constant cancellation rate and no diffusion, our model satisfactorily succeeds in replicating the dynamics of the empirical order density. This is largely attributed to the sophisticated submission rate of Eq. \eqref{eq:nu_r_param}. The presence of two different price scales for $x>0$ is of crucial importance. The time dependence of the first exponential term allows the order density around the auction price to increase as the clearing approaches. As the second exponential term is not time-dependent, the smooth price decay for larger $x$ is not disrupted (Fig. \ref{fig:num_sol_as_function_of_price}). This translates into an expansion of the order density only around the indicative price (Fig \ref{fig:num_sol_as_function_of_time}). Thus, agents posting orders near the indicative price can be seen as more sensitive to the auction deadline compared with agents that act at larger prices.

We note that the constant diffusion fit ($D_r,D_l>0$) permits the regularization of the obtained density around $x=0$ and $x=-x_0$ compared with the zero diffusion fit ($D_r=D_l=0$), where the discontinuity of the first derivative remains. Overall, the influence of diffusion is minimal within our framework, and adopting a time-varying revealed diffusion coefficient $D_r \propto 1/t$ yields slightly better fits around the indicative price. Lastly, locating the global minimum of the loss function $f$ given by Eq. \eqref{eq:obj_func} poses a considerable challenge and our strategy has been to seek approximate optimal parameters. We present these in Table \ref{tab:fitted_params}.
\begin{table}
\caption{Fitted parameters from the minimization of Eq. \eqref{eq:obj_func}. In the constant diffusion and the time diffusion cases, we fix the obtained parameters of the zero diffusion case.}
\begin{ruledtabular}
\begin{tabular}{lccc}
& Zero diffusion & Constant diffusion & Time diffusion \\
\hline
$C_r\cdot 10^{-1}$ & 9.3 & \_ & \_ \\
$x_r\cdot 10^{-3}$ & $2.3$ & \_ & \_ \\
$k\cdot 10^{0}$ & 4.9 & \_ & \_ \\
$w\cdot 10^{-1}$ & 8.7 & \_ & \_ \\
$\nu_l\cdot 10^{-2}$ & $2.3$ & \_ & \_ \\
$x_0\cdot 10^{-3}$ & $3.2$ & \_ & \_ \\
$m\cdot 10^{-2}$ & $1.6$ & \_ & \_ \\
$D_l\cdot 10^{-9}$ & 0 & 2.4 & 4.8 \\
$D_r\cdot 10^{-9}$ & 0 & 7 & \_ \\
$D_0\cdot 10^{-5}$ & \_ & \_ & $1.2$ \\
$D_T\cdot 10^{-8}$ & \_ & \_ & $2.2$ \\
\end{tabular}
\end{ruledtabular}
\label{tab:fitted_params}
\end{table}

\subsection{Discussion}

Even though the calibration procedure of section \ref{sec:num_solving} yields fairly good results, it is important to note that some simplifications were made. The null initial condition for the revealed order book $\rho^{r}(x,t=0)$ does not fully reflect reality, given that the order book is already populated by limit orders ($x>0$) prior to the closing auction. A large fraction of these orders are canceled at the start of the accumulation period, which results in an initial decrease in the revealed order book. This downward trend at $t=0$ is observable in Fig. \ref{fig:num_sol_as_function_of_time} for prices $0<x<1\%$. 

In addition, we have shown in section \ref{sec:mes_gamma_l} that cancellations exhibit a U-shaped pattern over time. We checked, however, that the introduction of a time-decreasing cancellation rate $\nu_l\Gamma_l$ analogous to Eq. \eqref{eq:D_r_param} does not significantly improve the quality of fits. Similarly, the time acceleration of cancellations around the indicative price does not enhance accuracy, given that the time acceleration of submissions can be adjusted to counterbalance it.

Note that plugging the empirical rates of submission and cancellation as measured in section \ref{sec:measures_1} into our model does not yield an accurate shape of the empirical order book. While the shape of empirical submissions is similar to the proposed functional of Eq. \eqref{eq:nu_r_param}, that of empirical cancellations seems to distort the order book at auction time. A likely reason for this discrepancy is the non-trivial behavior of high-frequency agents, which are responsible for intricate interactions and feedback loops that are not accounted for in our zero-intelligence framework.

A larger framework with fast and slow (or more) potentially interacting agents with markedly different submission, cancellation, and diffusion rates could be of interest but is beyond the scope of this paper. The total order density would then be a weighted sum of interacting individual order densities. Additionally, auctions are characterized by bursts of activity at specific round times typical of human behavior. These exogenous bursts of activity suggest segmenting the auction into distinct regimes (section \ref{sec:anomalous_scaling}). Lastly, our model does not capture the large peak of volumes at the indicative price: it is the result of strategic agents aiming to pin the auction price or simply sending orders at the current indicative price.

\section{The anomalous scaling of the indicative price}
\label{sec:anomalous_scaling}

When measuring the revealed diffusion coefficient $D_r$ in section \ref{sec:mes_diffusion}, the indicative price was assumed to be a diffusive process. However, it is known to be sub-diffusive \cite{challet2019strategic}. In this section, we examine the causes behind the anomalous diffusion of the indicative price.

\begin{figure}
    \centering
    \includegraphics[scale=0.36]{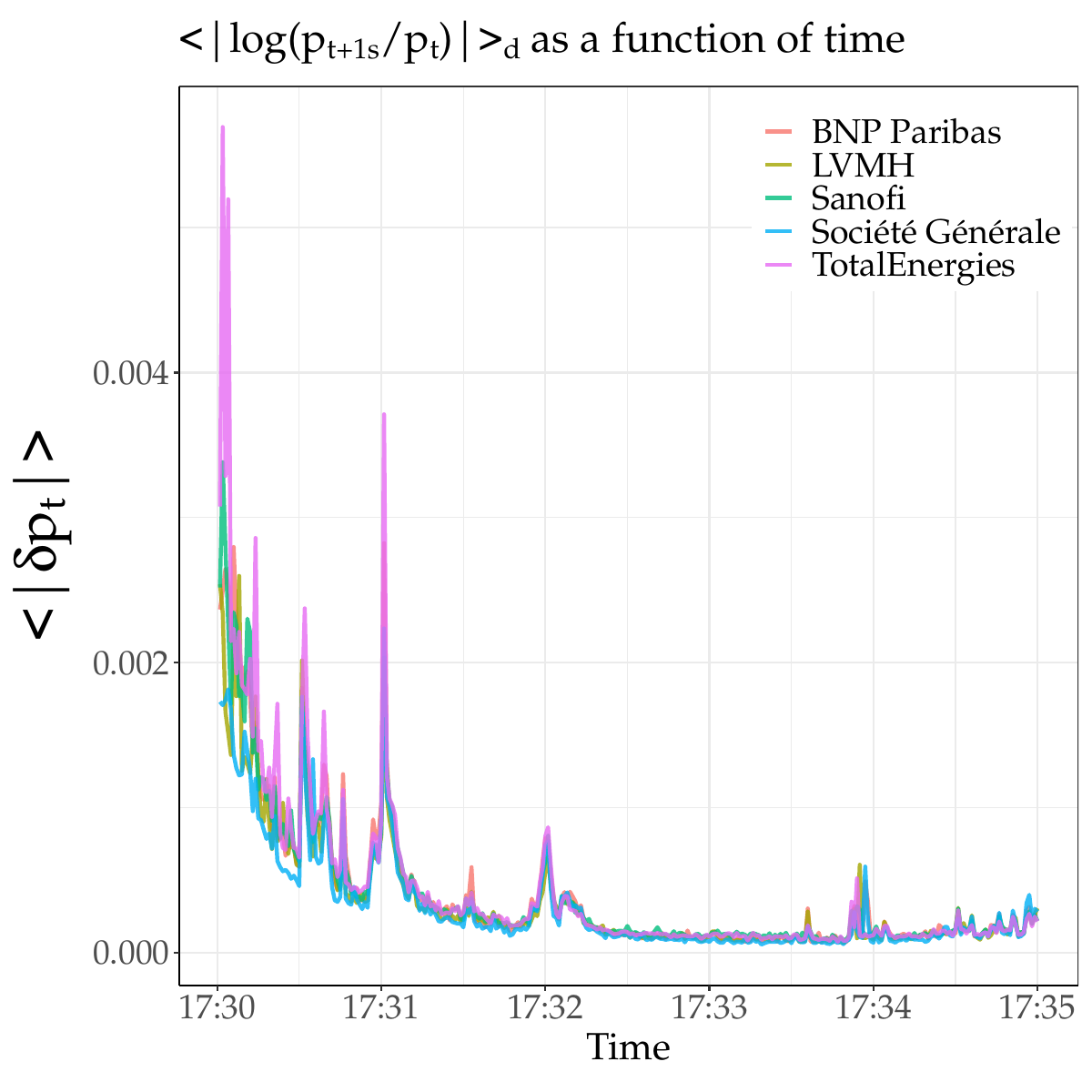}
    \includegraphics[scale=0.36]{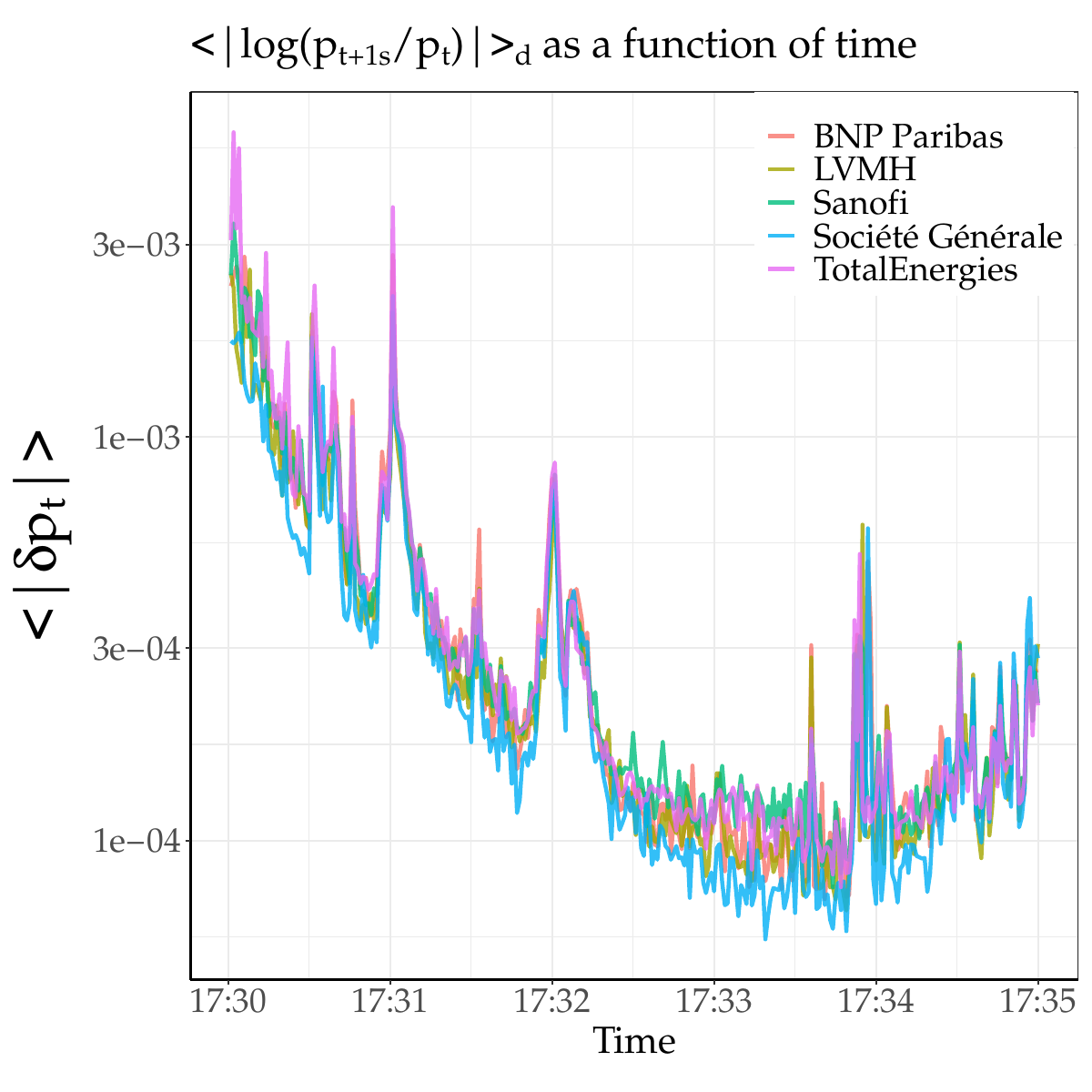}
    \caption{Ensemble average of the absolute value of 1-second log price increments $\mathbb{E}[|\log(p_{t+1\text{s}}/p_t)|]$ in the closing auction of five stocks traded on Euronext Paris between 2013 and 2017: BNP Paribas, LVMH, Sanofi, Société Générale, and TotalEnergies. Left panel: ordinary scale. Right panel: Y-axis in log scale.}
    \label{fig:E_p_t}
\end{figure}

We investigate the temporal pattern of the indicative price and find that it has non-stationary increments. Figure \ref{fig:E_p_t} depicts ensemble averages over days during the closing auction of 1-second absolute returns $\left<\left|\log(p_{t+1\text{s}}/p_t)\right|\right>$ for five stocks on Euronext Paris.
Here, we assume that the indicative price is a realization of the same stochastic process during each closing auction in order to proceed with ensemble averages for each stock. In particular, we observe volatility bursts and relaxations that suggest dividing the closing auction into different regimes: two 30-second regimes during the first minute, followed by a third regime between 17:31:00 and 17:32:00, a longer fourth regime between 17:32:00 and 17:34:00, and a final one-minute regime.

We test for the presence of anomalous scaling of the indicative price in each regime by computing average Hurst exponents. Assuming that the indicative price is a self-similar process, we can estimate an average Hurst exponent $H$ as 
\begin{equation}
    \left< \left< \left(\log(p_{t+\tau}/p_t)\right)^2\right>_t\right>_d \sim \tau^{2H},
\end{equation}
where $\left<\cdot\right>_t$ is the average over one realization of the indicative price series, and $\left<\cdot\right>_d$ is the ensemble average over days. The underlying process undergoes normal diffusion when $H=1/2$ and anomalous diffusion when $H\ne1/2$ (super-diffusion when $H>1/2$, and sub-diffusion when $H<1/2$).

\citet{chen2017anomalous} show that $H$ can be decomposed as $H=J+L+M-1$, where each of the exponents $J,L,M$ is associated with the failure of one condition of the central limit theorem: (i) the presence of correlations/long-term memory, (ii) infinite variance, and (iii) the non-stationarity of increments.
\begin{enumerate}
    \item Joseph exponent $J$ quantifies the long-term memory of increments. It is defined as the scaling of the ensemble average of a rescaled range statistic
    \begin{equation}
        \left<R_t/S_t\right>_d \sim t^J,
    \end{equation}
    where the considered range statistic is expressed as $R_t = \underset{1\leq s\leq t}{\max}\left[X_s - s/t X_t\right] - \underset{1\leq s\leq t}{\min}\left[X_s - s/t X_t\right]$, the deviation being $S_t^2 = Z_t/t - \left(X_t/t\right)^2$, and $X_t = \log(p_t/p_0)$ is the log indicative price centered around the origin of the considered regime. If $J>1/2$ increments are positively correlated, $J<1/2$ corresponds to negatively correlated increments, and $J=1/2$ refers to the absence of correlations; \citet{chen2017anomalous} argue that the Joseph exponent $J$ is the appropriate measure to test the efficient market hypothesis and not the Hurst exponent $H$.
    
    \item Noah exponent $L$ quantifies whether increments have finite variance ($L=1/2$), or infinite variance ($L>1/2$). Assuming that increments have power law tails with exponent $\gamma$, i.e. $\mathbb{P}_>(|x|) \underset{|x|\rightarrow +\infty}{\sim} |x|^{-\gamma}$, then 
    \begin{equation}
        L = \max(\frac{1}{2}, \frac{1}{\gamma}).
    \end{equation}
    
    \item Moses exponent $M$ quantifies whether increments are stationary ($M=1/2$) or non stationary ($M\ne1/2$). It is defined as the scaling of the ensemble average of centered absolute increments (see Fig. \ref{fig:E_p_t}) 
    \begin{equation}
    \left<|\log(p_{t+1}/p_t)|\right>_d \sim t^{M-\frac{1}{2}}.
    \end{equation}
\end{enumerate}

\begin{table}
\caption{Average Hurst, Joseph, Levy, and Moses exponents for the studied stocks in each regime.}
\label{tab:HJLM}
\begin{ruledtabular}
\begin{center}
\begin{tabular}{clccccr}
\multicolumn{1}{c}{Regime}&\multicolumn{1}{c}{Stock}&\multicolumn{1}{c}{H}&\multicolumn{1}{c}{J}&\multicolumn{1}{c}{L}&\multicolumn{1}{c}{M}&\multicolumn{1}{c}{H-J-L-M+1}\tabularnewline
\hline
&BNP Paribas&\_&0.47***&\_&\_&\_\tabularnewline
&LVMH&\_&0.56***&\_&\_&\_\tabularnewline
 1&Sanofi&\_&0.48***&\_&\_&\_\tabularnewline
&Société Générale&\_&0.41***&\_&\_&\_\tabularnewline
&TotalEnergies&\_&0.4***&\_&\_&\_\tabularnewline
\hline
&BNP Paribas&0.35***&0.54***&0.72***&0.09***&0\tabularnewline
&LVMH&0.43***&0.55***&0.71***&0.13***&0.04\tabularnewline
 2&Sanofi&0.38***&0.49&0.73***&0.13***&0.03\tabularnewline
&Société Générale&0.39***&0.49&0.78***&0.11***&0.01\tabularnewline
&TotalEnergies&0.38**&0.52.&0.69***&0.18***&-0.01\tabularnewline
\hline
&BNP Paribas&0.37***&0.63***&0.72***&0.03***&-0.01\tabularnewline
&LVMH&0.34***&0.54***&0.73***&0.06***&0.01\tabularnewline
 3&Sanofi&0.37***&0.52***&0.68***&0.1***&0.07\tabularnewline
&Société Générale&0.32***&0.53***&0.76***&-0.01***&0.04\tabularnewline
&TotalEnergies&0.34***&0.53***&0.71***&0.05***&0.04\tabularnewline
\hline
&BNP Paribas&0.35***&0.55***&0.73***&0.12***&-0.04\tabularnewline
&LVMH&0.36***&0.48***&0.69***&0.14***&0.05\tabularnewline
 4&Sanofi&0.34***&0.5&0.61***&0.26***&-0.03\tabularnewline
&Société Générale&0.37***&0.55***&0.74***&0.05***&0.03\tabularnewline
&TotalEnergies&0.36***&0.51**&0.68***&0.15***&0.01\tabularnewline
\hline
&BNP Paribas&0.58***&\_&\_&\_&\_\tabularnewline
&LVMH&0.55***&\_&\_&\_&\_\tabularnewline
 5&Sanofi&0.49&\_&\_&\_&\_\tabularnewline
&Société Générale&0.65***&\_&\_&\_&\_\tabularnewline
&TotalEnergies&0.51*&\_&\_&\_&\_\tabularnewline
\end{tabular}\end{center}
\end{ruledtabular}
  \begin{tablenotes}
   \item 
   \begin{enumerate}
    \item Values of $H$ in the first regime are omitted as the mean square displacement of the indicative price is not a power law of the lag in this regime and is rather noisy. We suspect the activation of VFA,VFC (Valid For Auction, Valid For Closing) to be the cause of this large noise.
    \item Values of $J$ in the last regime are omitted as the ensemble average of the rescaled range statistic is not a power law of time. Instead, $E[R_t/S_t]$ flattens with time: this results from the stabilization of the indicative price during the last moments of the accumulation period.
    \item    The symbols ***,**, and * indicate significance at the 0.1\%, 1\%, and 5\% level, respectively for testing the null hypothesis $\{S=0.5\}$, where $S \in \{H,J,L,M\}$.
   \end{enumerate}
  \end{tablenotes}
\end{table}

We estimate $H$, $J$, $L$, and $M$ for each regime separately. For $M$ and $L$, we use robust methods. Namely, $M$ is estimated using the median of the sum of absolute increments 
\begin{equation}
    m\left[\sum_{k=0}^{t-1} |\log(p_{k+1}/p_k)|\right] \sim t^{M+\frac{1}{2}},
\end{equation}
and $L$ is estimated using the median of the sum of square increments
\begin{equation}
    m\left[\sum_{k=0}^{t-1} \log(p_{k+1}/p_k)^2\right] \sim t^{2L+2M-1},
\end{equation}
where each sum starts at the beginning of the considered regime. For an in-depth discussion and proofs regarding the estimation of each exponent, see \citet{chen2017anomalous}. 

We report in Table \ref{tab:HJLM} numerical estimates of the scaling exponents in each regime for five different stocks. We draw the following remarks and conclusions from our results:
\begin{itemize}
    \item the indicative price is a sub-diffusive process during the closing auction as $H<1/2$, except in the final minute, where it switches to a diffusive (or an over-diffusive) behavior $H\gtrsim 1/2$; 
    \item the increments of the indicative price do not exhibit long-term memory during the accumulation period as $J\approx1/2$ for most stocks and regimes. Note that as the auction end approaches, the rescaled range statistic $R_t/S_t$ flattens due to the stabilization of the indicative price. As a result, $J$ is not defined in the last regime. We thus conclude that the indicative price is efficient in the sense of \citet{chen2017anomalous}: even if the apparent Hurst exponent is not $1/2$, $J=1/2$ precludes price predictions. Finally, \cite{euronext2021} report an overreaction of the indicative price based on the following definition of the indicative jump on close $J(t)$
\begin{equation}
    J(t) = \frac{p_t - p_{\text{ref}}}{p_{\text{auction}} - p_{\text{ref}}}.
\end{equation}
However, this incorporates the first jump from $p_{\text{ref}}$ (the last price of the continuous trading phase) to $p_0$ (the first indicative price of the accumulation period). When the reference price is the first indicative price of the accumulation period $p_0$, $J(t)$ becomes $\tilde{J}(t)$
\begin{equation}
    \tilde{J}(t) = \frac{p_t - p_0}{p_{\text{auction}} - p_0},
\end{equation}
which does not display any overreaction. 
In addition, we find that the $p_{\text{auction}}-p_{\text{ref}}$ does not have systematically the same sign as $p_{\text{auction}}-p_0$ (only $\approx$50\% of the time). Thus, the first jump from $p_{\text{ref}}$ to $p_0$ is not predictive of the direction of the auction price.
    \item the increments of the indicative price are highly non-stationary as $M$ significantly differs from $1/2$. Considering that $M\approx0$ implies that the indicative price volatility is a decreasing time function in each regime $\sigma_t \propto t^{-1/2}$, and equivalently that the revealed diffusion coefficient is time-dependent $D_r \propto 1/t$. This appears to be the major cause of the indicative price anomalous scaling;
    \item the increments of the indicative price exhibit infinite variance as $L\approx0.7$.
\end{itemize}

\section{Conclusion}

Zero intelligence models are surprisingly able to reproduce non-trivial stylized facts in financial markets \cite{farmer2005predictive}. Here, we showed that by adapting the zero intelligence latent/revealed liquidity framework of \citet{dall2019does} to auctions, we are able to replicate complex price-time dynamics of the average order book throughout the accumulation period. Within our framework, the skewed shape of the order book emerges from the product of the linear latent book by the exponentially decreasing submission rate. The time acceleration around the indicative price arises from inversely proportional rates to the remaining time to the deadline, analogously to typical human behavior when facing a deadline \cite{alfi2009people}. These results were confirmed by the estimation of the submission, cancellation, and diffusion rates. These represent a new piece of evidence advocating for the relevance of the latent order book of \citet{toth2011anomalous}. 

Although successful at reproducing many of the complex patterns observed during auctions, our model can only describe average order books where large daily fluctuations are neglected. Additionally, price-changing events (sell limit orders above the indicative price and buy limit orders below the indicative price that are breaching zero impacts) were supposed not to have an influence on the indicative price when they directly impact the supply/demand equilibrium similarly to market orders. Finite-size effects such as the large peak of volumes at the indicative price cannot be captured by continuous models and result from a possible strategic behavior. Finally, the heterogeneous nature of the agents involved and the reasons why they take part in auctions should be of interest: these range from manually trading agents for idiosyncratic reasons to more sophisticated trading algorithms minimizing impact and/or maximizing profits. A general model accounting for the overall auction ecology, price discreteness,  volume fluctuations, and the strategic behavior of agents is needed to explain daily deviations, e.g., during index rebalancing and derivatives expiry days, or after the release of a significant piece of news.

\section*{Acknowledgments}

We thank Michele Vodret for fruitful discussions and acknowledge the use of the EUROFIDAI BEDOFIH's database acquired through ``Equipex PLADIFES ANR-21-ESRE-0036 (France 2030)''.

\section*{Disclosure of interest}
The authors have no competing interests to declare.

\section*{Declaration of funding}
This publication stems from a partnership between CentraleSup\'elec and BNP Paribas.

\nocite{*}
\bibliography{apssamp}

\appendix

\section{Calibration of stationary order densities}
\label{sec:appendix_fits}
\begin{table}
\caption{Estimate values of the optimal parameters $(\nu_r/\nu_l \cdot a,\nu_r/\nu_l \cdot b,x_r,k,w)$ fitting the orders' density using the stationary setting. Fitting range: $x = 0 \pm 5\%$.}
\label{tab:static_fits_two_exp}
\begin{ruledtabular}
\begin{center}
\begin{tabular}{cccccccc}
\multicolumn{1}{c}{Side}&\multicolumn{1}{c}{Agent type}&\multicolumn{1}{c}{Stock}&\multicolumn{1}{c}{$\nu_r/\nu_l \cdot a$}&\multicolumn{1}{c}{$\nu_r/\nu_l \cdot b \cdot 10^{-2} $}&\multicolumn{1}{c}{$x_r$}&\multicolumn{1}{c}{$k$}&\multicolumn{1}{c}{$w$}\tabularnewline
\hline
&&BNP Paribas&5.72&0.56&0.37\%& 5.6& 0.984\tabularnewline
&&LVMH&9.23&0.62&0.3\%& 5.4& 0.978\tabularnewline
&ALL&Sanofi&7.89&0.58&0.29\%& 5.2& 0.983\tabularnewline
&&Société Générale&6.32&0.63&0.38\%& 6.1& 0.974\tabularnewline
&&TotalEnergies&6.08&0.50&0.33\%& 6.8& 0.989\tabularnewline
\cline{2-8}
&&BNP Paribas&5.24&0.47&0.36\%& 3.9& 0.980\tabularnewline
&&LVMH&7.47&0.53&0.31\%& 5.4& 0.986\tabularnewline
&MIX&Sanofi&5.89&0.47&0.31\%& 4.2& 0.986\tabularnewline
&&Société Générale&5.94&0.50&0.37\%& 6.8& 0.988\tabularnewline
Buy&&TotalEnergies&4.68&0.40&0.34\%& 5.0& 0.990\tabularnewline
\cline{2-8}
&&BNP Paribas&0.87&0.04&0.04\%& 7.8& 0.227\tabularnewline
&&LVMH&1.65&0.03&0.28\%& 5.6& 0.977\tabularnewline
&HFT&Sanofi&2.45&0.05&0.2\%&11.5& 0.995\tabularnewline
&&Société Générale&0.21&0.05&0.33\%& 3.3& 0.707\tabularnewline
&&TotalEnergies&1.91&0.04&0.21\%&10.5& 0.993\tabularnewline
\cline{2-8}
&&BNP Paribas&0.01&0.05&0.44\%& 6.8& 0.341\tabularnewline
&&LVMH&0.07&0.07&0.2\%&11.2& 0.591\tabularnewline
&NON&Sanofi&0.08&0.07&0.14\%&11.6& 0.631\tabularnewline
&&Société Générale&0.08&0.08&0.2\%&10.4& 0.317\tabularnewline
&&TotalEnergies&0.01&0.03&0.35\%& 7.6& \_ \tabularnewline
\hline
&&BNP Paribas&6.39&0.51&0.36\%& 7.4& 0.980\tabularnewline
&&LVMH&8.96&0.67&0.3\%& 4.7& 0.971\tabularnewline
&ALL&Sanofi&7.83&0.54&0.29\%& 5.4& 0.979\tabularnewline
&&Société Générale&6.83&0.60&0.37\%& 6.2& 0.977\tabularnewline
&&TotalEnergies&6.77&0.58&0.3\%& 5.1& 0.969\tabularnewline
\cline{2-8}
&&BNP Paribas&5.38&0.44&0.35\%& 3.7& 0.962\tabularnewline
&&LVMH&7.78&0.55&0.3\%& 5.3& 0.987\tabularnewline
&MIX&Sanofi&5.92&0.43&0.31\%& 4.8& 0.984\tabularnewline
&&Société Générale&6.23&0.48&0.35\%& 4.1& 0.976\tabularnewline
Sell&&TotalEnergies&5.12&0.46&0.32\%& 3.8& 0.971\tabularnewline
\cline{2-8}
&&BNP Paribas&1.00&0.02&0.29\%&12.1& 0.987\tabularnewline
&&LVMH&1.77&0.02&0.28\%& 7.2& 0.983\tabularnewline
&HFT&Sanofi&2.30&0.04&0.21\%&11.2& 0.995\tabularnewline
&&Société Générale&1.82&0.09&0.02\%&21.2& 0.585\tabularnewline
&&TotalEnergies&2.05&0.04&0.2\%& 7.7& 0.987\tabularnewline
\cline{2-8}
&&BNP Paribas&0.06&0.06&0.18\%&16.5& 0.452\tabularnewline
&&LVMH&0.08&0.06&0.27\%& 8.6& 0.516\tabularnewline
&NON&Sanofi&0.03&0.04&0.21\%& 9.6& 0.106\tabularnewline
&&Société Générale&0.10&0.07&0.44\%& 4.8& 0.291\tabularnewline
&&TotalEnergies&0.05&0.05&0.2\%&11.7& 0.258\tabularnewline
\end{tabular}\end{center}
\end{ruledtabular}
\end{table}

\end{document}